\documentclass[12pt]{article}

\usepackage{cite}
\usepackage{float} 
\usepackage{graphicx}
\usepackage{amsfonts}
\usepackage{float}
\usepackage{longtable}

\input{epsf}

\advance \topmargin by -\headheight
\advance \topmargin by -\headsep

\evensidemargin \oddsidemargin
\begin{document}

\def\rh{{\hat \rho}}
\def\alie{{\hat{\cal G}}}
\def\hpsi{{\widehat \psi}}
\newcommand{\sect}[1]{\setcounter{equation}{0}\section{#1}}
\renewcommand{\theequation}{\thesection.\arabic{equation}}

\def\rf#1{(\ref{eq:#1})}
\def\lab#1{\label{eq:#1}}
\def\nonu{\nonumber}
\def\br{\begin{eqnarray}}
\def\er{\end{eqnarray}}
\def\be{\begin{equation}}
\def\ee{\end{equation}}
\def\eq{\!\!\!\! &=& \!\!\!\! }
\def\foot#1{\footnotemark\footnotetext{#1}}
\def\lb{\lbrack}
\def\rb{\rbrack}
\def\llangle{\left\langle}
\def\rrangle{\right\rangle}
\def\blangle{\Bigl\langle}
\def\brangle{\Bigr\rangle}
\def\llbrack{\left\lbrack}
\def\rrbrack{\right\rbrack}
\def\lcurl{\left\{}
\def\rcurl{\right\}}
\def\({\left(}
\def\){\right)}
\newcommand{\nit}{\noindent}
\newcommand{\ct}[1]{\cite{#1}}
\newcommand{\bi}[1]{\bibitem{#1}}
\def\lskip{\vskip\baselineskip\vskip-\parskip\noindent}
\relax

\def\tr{\mathop{\rm tr}}
\def\Tr{\mathop{\rm Tr}}
\def\trace{\widehat{\rm Tr}}
\def\v{\vert}
\def\bv{\bigm\vert}
\def\Bgv{\;\Bigg\vert}
\def\bgv{\bigg\vert}
\newcommand\partder[2]{{{\partial {#1}}\over{\partial {#2}}}}
\newcommand\funcder[2]{{{\delta {#1}}\over{\delta {#2}}}}
\newcommand\Bil[2]{\Bigl\langle {#1} \Bigg\vert {#2} \Bigr\rangle}  
\newcommand\bil[2]{\left\langle {#1} \bigg\vert {#2} \right\rangle} 
\newcommand\me[2]{\left\langle {#1}\bv {#2} \right\rangle} 
\newcommand\sbr[2]{\left\lbrack\,{#1}\, ,\,{#2}\,\right\rbrack}
\newcommand\pbr[2]{\{\,{#1}\, ,\,{#2}\,\}}
\newcommand\pbbr[2]{\lcurl\,{#1}\, ,\,{#2}\,\rcurl}

\def\ket#1{\mid {#1} \rangle}
\def\bra#1{\langle {#1} \mid}
\newcommand{\braket}[2]{\langle {#1} \mid {#2}\rangle}
%
\def\a{\alpha}
\def\at{{\tilde A}^R}
\def\atc{{\tilde {\cal A}}^R}
\def\atcm#1{{\tilde {\cal A}}^{(R,#1)}}
\def\b{\beta}
\def\dc{{\cal D}}
\def\d{\delta}
\def\D{\Delta}
\def\eps{\epsilon}
\def\vareps{\varepsilon}
\def\g{\gamma}
\def\G{\Gamma}
\def\grad{\nabla}
\def\h{{1\over 2}}
\def\l{\lambda}
\def\L{\Lambda}
\def\m{\mu}
\def\n{\nu}
\def\o{\over}
\def\om{\omega}
\def\O{\Omega}
\def\p{\phi}
\def\P{\Phi}
\def\pa{\partial}
\def\pr{\prime}
\def\pt{{\tilde \Phi}}
\def\qs{Q_{\bf s}}
\def\ra{\rightarrow}
\def\s{\sigma}
\def\S{\Sigma}
\def\t{\tau}
\def\th{\theta}
\def\Th{\Theta}
\def\tpp{\Theta_{+}}
\def\tmm{\Theta_{-}}
\def\tpg{\Theta_{+}^{>}}
\def\tms{\Theta_{-}^{<}}
\def\tp0{\Theta_{+}^{(0)}}
\def\tm0{\Theta_{-}^{(0)}}
\def\ti{\tilde}
\def\wti{\widetilde}
\def\jc{J^C}
\def\bj{{\bar J}}
\def\sj{{\jmath}}
\def\bsj{{\bar \jmath}}
\def\bp{{\bar \p}}
\def\vp{\varphi}
\def\ve{\varepsilon}
\def\vt{{\tilde \varphi}}
\def\faa{Fa\'a di Bruno~}
\def\ca{{\cal A}}
\def\cb{{\cal B}}
\def\ce{{\cal E}}
\def\cg{{\cal G}}
\def\cgh{{\hat {\cal G}}}
\def\ch{{\cal H}}
\def\chh{{\hat {\cal H}}}
\def\cl{{\cal L}}
\def\cm{{\cal M}}
\def\cn{{\cal N}}
\def\u2{\mid u\mid^2}
\def\ub{{\bar u}}
\def\z2{\mid z\mid^2}
\def\zb{{\bar z}}
\def\w2{\mid w\mid^2}
\def\wb{{\bar w}}
\newcommand\sumi[1]{\sum_{#1}^{\infty}}   
\newcommand\fourmat[4]{\left(\begin{array}{cc}  
{#1} & {#2} \\ {#3} & {#4} \end{array} \right)}

%
\def\lie{{\cal G}}
\def\kmlie{{\hat{\cal G}}}
\def\dlie{{\cal G}^{\ast}}
\def\elie{{\widetilde \lie}}
\def\edlie{{\elie}^{\ast}}
\def\hlie{{\cal H}}
\def\flie{{\cal F}}
\def\wlie{{\widetilde \lie}}
\def\f#1#2#3 {f^{#1#2}_{#3}}
\def\winf{{\sf w_\infty}}
\def\win1{{\sf w_{1+\infty}}}
\def\hwinf{{\sf {\hat w}_{\infty}}}
\def\Winf{{\sf W_\infty}}
\def\Win1{{\sf W_{1+\infty}}}
\def\hWinf{{\sf {\hat W}_{\infty}}}
\def\Rm#1#2{r(\vec{#1},\vec{#2})}          
\def\OR#1{{\cal O}(R_{#1})}           
\def\ORti{{\cal O}({\widetilde R})}           
\def\AdR#1{Ad_{R_{#1}}}              
\def\dAdR#1{Ad_{R_{#1}^{\ast}}}      
\def\adR#1{ad_{R_{#1}^{\ast}}}       
\def\KP{${\rm \, KP\,}$}                 
\def\KPl{${\rm \,KP}_{\ell}\,$}         
\def\KPo{${\rm \,KP}_{\ell = 0}\,$}         
\def\mKPa{${\rm \,KP}_{\ell = 1}\,$}    
\def\mKPb{${\rm \,KP}_{\ell = 2}\,$}    
%
\def\rlx{\relax\leavevmode}
\def\inbar{\vrule height1.5ex width.4pt depth0pt}
\def\IZ{\rlx\hbox{\sf Z\kern-.4em Z}}
\def\IR{\rlx\hbox{\rm I\kern-.18em R}}
\def\IC{\rlx\hbox{\,$\inbar\kern-.3em{\rm C}$}}
\def\IN{\rlx\hbox{\rm I\kern-.18em N}}
\def\IO{\rlx\hbox{\,$\inbar\kern-.3em{\rm O}$}}
\def\IP{\rlx\hbox{\rm I\kern-.18em P}}
\def\IQ{\rlx\hbox{\,$\inbar\kern-.3em{\rm Q}$}}
\def\IF{\rlx\hbox{\rm I\kern-.18em F}}
\def\IG{\rlx\hbox{\,$\inbar\kern-.3em{\rm G}$}}
\def\IH{\rlx\hbox{\rm I\kern-.18em H}}
\def\II{\rlx\hbox{\rm I\kern-.18em I}}
\def\IK{\rlx\hbox{\rm I\kern-.18em K}}
\def\IL{\rlx\hbox{\rm I\kern-.18em L}}
\def\one{\hbox{{1}\kern-.25em\hbox{l}}}
\def\0#1{\relax\ifmmode\mathaccent"7017{#1}%
B        \else\accent23#1\relax\fi}
\def\omz{\0 \omega}
%
\def\ltimes{\mathrel{\vrule height1ex}\joinrel\mathrel\times}
\def\rtimes{\mathrel\times\joinrel\mathrel{\vrule height1ex}}
%
\def\mark{\noindent{\bf Remark.}\quad}
\def\prop{\noindent{\bf Proposition.}\quad}
\def\theor{\noindent{\bf Theorem.}\quad}
\def\name{\noindent{\bf Definition.}\quad}
\def\exam{\noindent{\bf Example.}\quad}
\def\proof{\noindent{\bf Proof.}\quad}

\begin{titlepage}
\vspace*{-1cm}

\vskip 3cm

\vspace{.2in}
\begin{center}
{\large\bf  A False Vacuum Skyrme Model for Nuclear Matter }
\end{center}

\vspace{.5cm}

\begin{center}
L. A. Ferreira$^{\dagger ,}$\footnote{laf@ifsc.usp.br} and L. R. Livramento$^{\dagger ,\star ,}$\footnote{livramento@theor.jinr.ru}

\vspace{.3 in}
\small

\par \vskip .2in \noindent
$^{\dagger}$Instituto de F\'\i sica de S\~ao Carlos; IFSC/USP;\\
Universidade de S\~ao Paulo, USP  \\ 
Caixa Postal 369, CEP 13560-970, S\~ao Carlos-SP, Brazil\\

\par \vskip .2in \noindent
$^{\star}$BLTP, JINR, Dubna 141980, Moscow Region, Russia

\vskip 2cm

\begin{abstract}

The low energy regime of Quantum Chromodynamics (QCD) presents enormous challenges due to its large coupling. Effective field theories, like the Skyrme model, are useful approaches to study properties of strong interaction at hadronic scales.  We propose a  Skyrme-type  model  with  a self-dual sector and that treats the density of the baryonic charge as a self-interacting fluid. The dynamics reduces to Coleman's false vacuum problem for a scalar field that is a fractional power of that density. The main result is that such a Skyrme-type model is the first one  to reproduce, with good accuracy, the experimental values of radii and binding energies  for a very wide range of the mass number. The robust and simple properties of the model lead to many possible generalizations with implications not only in nuclear physics but also in other areas of Physics.

\end{abstract}

\normalsize
\end{center}
\end{titlepage}

\section{Introduction}
\label{sec:introduction}
\setcounter{equation}{0}

QCD is a quite successful  non-abelian gauge theory for the strong interactions of quarks and gluons.  It presents asymptotic freedom and it is in very good agreement with experiment at high energies. However, the dynamics of QCD at hadronic scales presents enormous challenges due to its large coupling, and the developments of non-perturbative methods are still well below the needs for an understanding of   QCD low energy physics.  In order to deal with the massive amount of experimental data,  physicists have used all the resources available to construct models for the strong-coupled dynamics among baryons, not necessarily derived from QCD.  Nowadays, the state of the art includes, besides lattice gauge theory, {\em ab initio} models like the no core shell model \cite{vary},  truncations of it \cite{stetcu}, high-low momentum decoupling models \cite{furnstahl}, as well as low energy effective theories like the inspiring {\em chiral perturbation theory}  put forward by S. Weinberg \cite{weinberg,ordonez,vankolck}. On the other hand, some important studies on non-abelian gauge theories have shed light on particular sectors of strong-coupled QCD.  It has been shown  that in the large N-limit, QCD may be equivalent to a theory of mesons, and  baryons could be identified with the Skyrmions \cite{thooft,witten1,witten2,adkins}. 

That observation boosted the interest in an old idea of T. Skyrme, where baryons and  nuclei can be interpreted  as topological solitons of  the so-called Skyrme model \cite{skyrme1,skyrme2}. Skyrmion solutions with various baryonic topological charges have been constructed using powerful numerical techniques  \cite{sut1,sut2,mantonbook}. However, the binding energies obtained are too high when compared with  experiment, and that is perhaps linked with the fact that the original Skyrme model does not admit a self-dual sector  \cite{mantonruback,derek}.  Modifications of the Skyrme model  leading to  self-dual sectors have made some improvement on the problem of the binding energies. 

One such attempt  reduces self-dual Yang-Mills  in four dimensions to a Skyrme model coupled to an infinite tower of vector mesons, with some success  for the spectrum of light nuclei \cite{sut_tower,sut_naya_1,sut_naya_2}. Another attempt  deals with modifications of the so-called BPS Skyrme model \cite{adam1,adam2} that has obtained reasonable values of binding energies for very heavy nuclei and on some aspects of neutron stars \cite{adam_prl,adam_neutron_star}. The Skyrme model has been applied to other aspects of nuclear matter \cite{skyrmereview,skyrmebook}, and recently, interesting results were obtained on the spectra of light nuclei \cite{mantonmanko,mantonwood,halcrow1,halcrow2} and on the spin-orbit interaction \cite{spinorbit}. 

In this paper we report results for the values of radii and binding energies for Skyrmions with topological charges varying from 1 to 240, obtained in a modification of the self-dual Skyrme model proposed in \cite{laf2017}.  They  are in quite good agreement with the experimental values of root-mean-square (rms) charge radii \cite{angeli} and binding energies per nucleon \cite{ame2016}, in a list of $N_c=265$ nuclei, containing all the  stable nuclei up to $^{208}$Pb, and above that  with those of nuclei up to $^{240}$Pu with a half-life greater than $10^3$ years, according to \cite{nubase2016}.  It is the first Skyrme-type model to reproduce those experimental results with a good accuracy and for such a wide range of mass number. 

The model that we present is made of two parts. The first one is the self-dual Skyrme model proposed in \cite{laf2017} and further explored in \cite{us}, and it consists of a modification of the original  Skyrme model \cite{skyrme1,skyrme2} where the group indices in the action are contracted, not with the Killing form of $SU(2)$, but with a symmetric matrix $h$ of scalar fields in the quadratic term of the action, and its inverse in the quartic term. It presents an exact self-dual sector and it is conformally invariant in three space dimensions. The second part is a theory of a self-interacting fluid where the role of the order parameter, denoted $\psi$, is played by a fractional power of the density of the baryonic (topological ) charge. The two parts are coupled through the chiral $SU(2)$ Skyrme fields as the fluid theory does not involve the scalar fields in the matrix $h$. An interesting aspect of the model is that the $h$-fields adjust themselves to solve the self-duality equations of the first part of the model, and consequently the $SU(2)$ Skyrme fields are not constrained by the equations of motion coming from that first part. The dynamics of the $SU(2)$ Skyrme fields are determined solely by the equations of the fluid theory.   

As a  consequence of such interesting coupling, the contribution to the energy of the first part  is proportional to the topological charge, as it satisfies the self-duality equations.  As such topological charge is the baryonic charge, and so equal to the number of nucleons, it turns out  that the energy of the first part gives the masses of the uncoupled nucleons. The contribution to the energy of the self-interacting fluid is responsible for the binding energy of the nucleons inside the nuclei. The two contributions have different energy scales. Such a two scale regime is a desired feature, since it is expected from QCD that nucleons are bound states of quarks and gluons and nuclei are bound states of nucleons, and they involve different energy scales. 

The fluid theory have kinetic and potential terms for the order parameter $\psi$. Since the density of the topological charge involve first derivatives of the $SU(2)$ Skyrme fields, the action of the fluid theory contains second derivatives of those fields. However, due to the topological character of the order parameter $\psi$, the Euler-Lagrange equations for the $SU(2)$ Skyrme fields, associated to the fluid theory, lead to a second order partial differential equation for $\psi$. The effect of the third derivative is to introduce an integration constant on those equations, which  plays a crucial role in the properties of the model. All the physical properties of the model depend upon such a constant. In particular, the topological charge turns out to be a monotonically decreasing function of it. Such an  integration constant plays a role similar to the running coupling constants in the renormalization group. Indeed, we have to choose its value for a reference nucleus  in order to fix the value of the physical quantities (mass, baryonic charge, radius, etc) for all other nuclei. 

Another interesting feature of the model is that the dynamics of the self-interacting fluid theory reduces to the false vacuum problem of S. Coleman \cite{coleman1,coleman2}. It follows that  the solution for the order parameter $\psi$ that presents the lowest energy, for a given value of the  baryonic charge, is the Coleman's bounce solution and so it is spherically symmetric and decays monotonically (exponentially) to zero at large distances. In addition, as the integration constant decreases, and so the baryonic charge increases, the Coleman's false vacuum tend to become degenerate with the true vacuum. A consequence of that is that the order parameter $\psi$ remains practically constant for a large region around the origin, before decaying exponentially to zero at larger distances.  Therefore, the heavier the nucleus is, the smaller the value of the corresponding  integration constant is, and so the closer it is to that regime. Consequently, such a behavior, together with the spherical symmetry of the solutions, leads in quite natural and robust way to the proportionality between the root mean square (RMS) radius of the nucleus and  its baryonic charge risen to the power one third. 
 
 The properties of the Coleman's false vacuum problem also lead to a robust behaviour of the binding energy. We are able to show, for a quite large part of the parameter space of the model, that the binding energy per nucleon, provided by Coleman's bounce solutions, grows with the baryonic charge and saturates for large values of it. However, the energy functional of the second part of our model possess a topological term which does not affect the equations of motion of the fluid theory. That term can be chosen to be proportional to the square of the baryonic charge and so, for the type of nuclei we consider, it resembles the Coulomb interaction energy among the protons. Therefore, such a topological term  will make the binding energy to decrease for large values of the baryonic charge. Consequently, our model reproduces the bulk behaviour of the binding energy per nucleon of a large class of nuclei, for a large portion of parameter space. Those parameters have to be fixed just to fit the scales of the problem, leading to a very good agreement with the experimental data, with errors smaller than $1\%$ for baryonic charges running from 20 to 240, and with even smaller errors for baryonic charges above 60. The error grows a bit for nuclei with baryonic charges between 10 and 20, but not higher than $5\%$. Only for very light nuclei, with baryonic charges below 10, the error grows to a two digit percentage.
 
The model describes heavier nuclei better than light ones because of the spherical symmetry of the solutions, imposed by Coleman's false vacuum problem. In order to improve the description of light nuclei we have perhaps to add kinetic and potentials terms for the scalar fields associated to the symmetric matrix $h$, in order to break the self-duality. Such a breaking has to be small as the deviations we have from the experimental data are already small. So,  such a breaking  of self-duality can perhaps be analysed through a perturbative expansion on the couplings constants of such new terms involving the $h$-fields. In addition, we expect that the breaking of the self-duality will bring the properties of the first part of the model closer to those of the original Skyrme model. Therefore, many results already known for the Skyrme model can perhaps be analysed in the context of the present model.

The paper is organised as follows. In section \ref{sec:model} we introduce our  model, made of the self-dual Skyrme model, introduced in \cite{laf2017,us}, coupled to a self-interacting fluid  with an order parameter which is a fractional power of the density of baryonic charge.  In section \ref{sec:falsevacuum} we show that the dynamics of the fluid theory reduces to the Coleman's false vacuum problem \cite{coleman1,coleman2} and discuss its consequences. The analysis of the bulk properties of the binding energy per nucleon is made in section \ref{sec:binding}. The choice of an admissible potential, in the sense of \cite{coleman1,coleman2}, is made in section \ref{sec:potential}, and it is shown that the model presents a robust behaviour that agrees with the bulk properties of nuclei. The analysis of the numerical and experimental data is made in section \ref{sec:data}, and our conclusions are presented in section \ref{sec:conclusion}. In the Appendix  \ref{sec:numerical} we describe our numerical methods, and in the Appendix \ref{sec:longtable} we provide a long table with the detailed numerical and experimental data used in the analysis of section \ref{sec:data}.

\section{The  model}
\label{sec:model}
\setcounter{equation}{0}

The first part of our model consists of a modified Skyrme model \cite{laf2017,us} containing,  besides the usual chiral fields $U\in SU(2)$, six scalar fields assembled in a $3\times 3$, symmetric and invertible  matrix $h_{ab}$, with all its eigenvalues being positive, and defined by the static energy 
\be
E_1= \int d^3x\left[ \frac{m_0^2}{2}\, h_{ab}\,R^a_{i}\,R^{b}_{i}+\frac{1}{4\,e_0^2}\, h^{-1}_{ab}\,H^a_{ij}\,H^{b}_{ij}\right]
\lab{e1}
\ee
where $m_0$ and $e_0$ are coupling constant with dimension of $[m_0]=\sqrt{[{\rm energy}]/[{\rm length}]}$ and $[e_0]=1/\sqrt{[{\rm energy}][{\rm length}]}$. In addition,  
\be
R^a_{\mu}=i\, \trace\(\partial_{\mu}U\,U^{\dagger}\,T_a\);\qquad\qquad \qquad 
H^a_{\mu\nu}=\ve_{abc}\,R_{\mu}^b\,R_{\nu}^c
\lab{rhdef}
\ee 
with $T_a$, $a=1,2,3$, being the basis of the $SU(2)$ Lie algebra satisfying $\sbr{T_a}{T_b}=i\,\ve_{abc}\,T_c $, and 
$\trace\(T_a\,T_b\)=\delta_{ab}$, is a normalised trace. In a given representation of $SU(2)$ we have that ${\rm Tr}\(T_a\,T_b\)=k\,\delta_{ab}$, with $k$ being proportional to the Dynkin index of that representation. We define $\trace\equiv \frac{1}{k}\,{\rm Tr}$. Following  the ideas of  \cite{genbps}, the introduction of  $h_{ab}$  makes  possible  the existence of a self-dual sector \cite{bpswojtek,bpsshnir}. The self-duality equations are given by 
\be
\lambda \,h_{ab}\,R^b_i=\frac{1}{2}\,\ve_{ijk}\,H^a_{jk}\qquad\qquad {\rm with} \qquad \qquad\lambda=\pm\,m_0\,e_0
\lab{selfdual}
\ee 
The sign of $\lambda$ characterises the self-dual and anti-self-dual sectors of  \rf{e1}, and the solutions on those sectors saturate a bound on the energy, such that  
\be 
E_{1}=48\,\pi^2\,\frac{m_0}{e_0}\,\mid Q\mid 
\lab{e1top}
\ee 
where $Q$ is the topological charge of the Skyrme model, i.e. 
\be Q= \frac{i}{48\,\pi^2}\,\int d^3x\; \ve_{ijk}\,\trace\(R_i\,R_j\,R_k\) \lab{charge0}\ee
 with $R_i=R_i^a\,T_a$.  It follows from self-duality that
 \be
 {\rm sign}\,\(\lambda\,Q\)=-1
 \lab{signlambdaq}
 \ee
 The introduction of  $h_{ab}$ also renders the theory \rf{e1} conformally invariant in the three dimensional space \cite{laf2017,us}. 

The results of \cite{us} relevant for the construction of the model proposed in this paper  are: {\em i)} the first order self-duality equations \rf{selfdual} imply the nine static second order Euler-Lagrange equations associated  to  fields $U$ and $h_{ab}$, {\em ii)}  the static Euler-Lagrange equations associated to the fields $h_{ab}$ are equivalent to the self-duality equations, {\em iii)}  given a configuration for the $U$-fields one can solve the self-duality equations by taking $h_{ab}$ to be
\be
h=\frac{\sqrt{{\rm det}\,\tau}}{ m_0\,e_0}\; \tau^{-1};\qquad\qquad \tau_{ab}= R_{i}^a\,R_{i}^b
\ee 
So, the fields $h_{ab}$ are spectators in the sense that they adjust themselves to solve the self-duality equations for any configuration of the $U$-fields. Note that the matrix $\tau$ is similar to the Skyrme model strain tensor  \cite{mantonbook}. For  $U$-field configurations where $\tau$ is singular the matrix $h_{ab}$ still solves the self-duality equation but it is not completely determined by $U$, and have some arbitrary components \cite{us}. 

Therefore, if one modifies the theory \rf{e1} by adding a  static energy $E_2$ that depends only on the $U$-fields, the Euler-Lagrange equations associated to the $h$-fields will not change and so the self-duality equations will continue to be valid, and consequently the matrix $h$ will be determined from the $U$-fields. In addition, the part of the Euler-Lagrange equations associated to the $U$-fields coming from $E_1$ will be automatically satisfied, since the self-duality equations still hold true. So, the configurations of the $U$-fields  will only be subjected to the $E_2$-part of their Euler-Lagrange equations. Consequently, the addition of such $U$-dependent $E_2$, may break the conformal symmetry of the theory \rf{e1}, but will keep its self-dual sectors intact. So, using \rf{e1top}, the static energy of our model is given by 
\be 
E=48\,\pi^2\,\frac{m_0}{e_0}\,\mid Q\mid +E_2
\lab{tot}
\ee
and we shall take $E_2$ to be a fluid theory for a quantity $\psi$ which is  proportional to a fractional power of the  density of the topological charge, i.e. 
\br
\psi^s\equiv -\frac{i}{12\,\lambda^3}\,\ve_{ijk}\,{\widehat{\rm Tr}}\(R_i\,R_j\,R_k\)
\lab{psidef}
\er
 with $s$ being a real positive  parameter. The  introduction of  $\lambda$ in \rf{psidef} makes $\psi$  to be dimensionless and positive. The self-duality equation implies that  ${\rm det}\,h=4\,\psi^s$ \cite{us}, and so the positivity of $\psi$ follows from the fact that the eigenvalues of $h$ are positive in order for  $E_1$ to be positive. We take $E_2$ to be
\br
E_2=\int d^3x\;\left[\frac{\mu_0^2}{2}\,\(\partial_i\psi\)^2+V\(\psi\) + G\(U\)\,\psi^s\right]
\lab{e2}
\er
where $\mu_0$ is a coupling constant with dimension of mass, and $G\(U\)$ is a functional of the   $U$-fields  but not of their derivatives, which we shall take to be positive. Note that the $G$-term in \rf{e2} is topological in the sense that it is invariant under any smooth variation of the $U$-fields. It differs from the Skyrme topological charge $Q$ since $G$ works like a deformation of  the target space and may break the $SU(2)_L\otimes SU(2)_R$ symmetry of  \rf{e1}.  Since $\psi$ depends upon the first derivatives of the $U$-fields, it follows that $E_2$ presents second derivatives of those fields. However, the Euler-Lagrange equations for the $U$-fields take a simple form in terms of $\psi$. In order to see that, it is easier to use  Darboux-like coordinates on $SU(2)$ given by
\be
U=\(\begin{array}{cc}
\sqrt{1-F}\,e^{i\,\theta_2}& i\,\sqrt{F}\,e^{i\,\theta_1}\\
i\,\sqrt{F}\,e^{-i\,\theta_1}&\sqrt{1-F}\,e^{-i\,\theta_2}
\end{array}\);\qquad\qquad
0\leq F\leq 1;\qquad 0\leq \theta_1\,,\,\theta_2\leq 2\,\pi
\lab{darbouxsu2}
\ee
and \rf{psidef} becomes 
\be
\psi^s=\frac{1}{\lambda^3}\,\ve_{ijk}\,\partial_i F\,\partial_j\theta_1\,\partial_k\theta_2
\ee
Then the Euler-Lagrange equations following from \rf{e2}, associated to the fields $F$, $\theta_1$ and $\theta_2$ are, respectively,
\br
\ve_{ijk}\,\partial_j\theta_1\,\partial_k\,\theta_2\,\partial_i\left[\psi^{1-s}\(-\mu_0^2\, \partial_l^2\psi+\frac{\delta \,V}{\delta\,\psi}\)\right]&=&0
\nonumber\\
\ve_{ijk}\,\partial_i F\,\partial_k\,\theta_2\,\partial_j\left[\psi^{1-s}\(-\mu_0^2\, \partial_l^2\psi+\frac{\delta \,V}{\delta\,\psi}\)\right]&=&0
\lab{darbouxeqs}\\
\ve_{ijk}\,\partial_iF\,\partial_j\theta_1\,\partial_k\,\left[\psi^{1-s}\(-\mu_0^2\, \partial_l^2\psi+\frac{\delta \,V}{\delta\,\psi}\)\right]&=&0
\nonumber
\er
Therefore, it turns out that the three Euler-Lagrange associated to the $U$-fields are equivalent to 
\br
\mu_0^2\,\partial^2 \psi -\frac{\delta\,V_{\rm eff.}}{\delta\,\psi}=0\;;\qquad\qquad V_{\rm eff.}\equiv V-c\, \psi^{s}
\lab{psieq}
\er
where $c$ is an arbitrary integration constant. Clearly, if we take $\psi$ as the fundamental degree of freedom, instead of the $U$-fields, \rf{psieq} is the Euler-Lagrange equation following from the functional 
\br
E_{\rm eff.}=\int d^3x\;\left[\frac{\mu_0^2}{2}\,\(\partial_i\psi\)^2+V\(\psi\) - c\,\psi^s\right]
\lab{eeff}
\er

The equation \rf{psieq} is the only one we have to solve to construct the Skyrmions, and  for each choice of $c$ there will be at least one solution. The subtlety however is that the acceptable solutions  are those such that  the Skyrme topological charge 
\be 
Q=-\frac{\lambda^3}{4\,\pi^2}\, \int d^3x\, \psi^s 
\lab{top0}
\ee 
is an integer, otherwise we will not be able to integrate  \rf{psidef}  to obtain the solution for the $U$-fields with the appropriate boundary conditions. 
The parameter $c$ therefore plays the role of a running coupling constant, and we will have to choose which value of it corresponds to a chosen value of the topological charge.  Once that choice is made the allowed values of $c$ will fall into a discrete sequence of real numbers. 

Note that from \rf{signlambdaq} it follows that
\be 
\mid Q\mid =\frac{\mid\lambda\mid^3}{4\,\pi^2}\, \int d^3x\, \psi^s 
\lab{topmodulus}
\ee 

\section{The false vacuum}
\label{sec:falsevacuum}
\setcounter{equation}{0}

Obviously, the density of baryonic charge must go to zero at infinity, and so we need $\psi\rightarrow 0$ as $r\rightarrow \infty$. Therefore, $\psi=0$ must be an extrema of $ V_{\rm eff.}$. In addition, in order to have $E_{\rm eff.}$ finite we need $ V_{\rm eff.}$ to vanish at $\psi=0$. So we need
\be
\psi\mid_{r\rightarrow \infty}=0;\qquad \quad V_{\rm eff.}\mid_{\psi=0}=0;\qquad \quad \frac{\delta\,V_{\rm eff.}}{\delta\,\psi}\mid_{\psi=0}=0
\lab{effpotcond}
\ee
A quite robust property of nuclei is that the nuclear matter density $\rho$ falls exponentially at large distances as $\rho\(r\)\sim \rho_0\,e^{-r/a}$, with $a$ being roughly independent of the mass number and given approximately by $a=0.524$ Fermi. Our model does have the property that all solutions, with any value of the baryonic charge, fall off exponentially at the same rate. From \rf{psidef} we must identify $\rho$ with $\psi^s$. Therefore, expanding \rf{psieq} around $\psi=0$, one observes that $\psi$ behaves as
\be
 \psi \sim \frac{e^{-r/(s\,a)}}{r};\qquad\qquad  {\rm for}\qquad  r\rightarrow \infty
 \lab{psiyukawa}
 \ee
if we assume that 
\br
 \frac{\delta^2\,V_{\rm eff.}}{\delta\,\psi^2}\mid_{\psi=0}=\(\frac{\mu_0}{s\,a}\)^2
\lab{massscale}
\er
So, $\psi=0$ must be  a local minimum of $V_{\rm eff.}$. 

The solutions of \rf{psieq} must be stable under  Derrick's argument \cite{derrick,coleman2}. Indeed, writing \rf{eeff} as 
\be
E_{\rm eff.}= T+U_{\rm eff.}; \qquad  T= \frac{\mu_0^2}{2}\,\int d^3x\,\(\partial_i\psi\)^2;\qquad U_{\rm eff.}=\int d^3x\;V_{\rm eff.}
\ee
and defining the scale transformation $\psi_{\alpha}\(x\)=\psi\(x/\alpha\)$, we have that $T\rightarrow\alpha\,T$ and $U_{\rm eff.}\rightarrow\alpha^3\,U_{\rm eff.}$. Any solution of \rf{psieq} makes $E_{\rm eff.}$ stationary, and so $E_{\rm eff.}$ should be stationary under scale transformation when \rf{psieq} is imposed. Therefore, we must have 
\be 
T+3\,U_{\rm eff.}=0 ;\qquad\qquad{\rm and\:\, so}\qquad\qquad E_{\rm eff.}=\frac{2}{3}\,T
\lab{derrick0}
\ee 
 Thus, $E_{\rm eff.}$ is positive and $U_{\rm eff.}$ is negative when evaluated on solutions of \rf{psieq}. That means $V_{\rm eff.}$ must be negative for some region of values of $\psi$. We are then led to  S. Coleman's  false vacuum problem \cite{coleman1,coleman2}.  The main result of \cite{coleman2} is  that, for what it defines as an admissible effective potential $V_{\rm eff.}$, the equation \rf{psieq} possesses at least one monotone spherically symmetric solution  $\psi_c$ vanishing at infinity, other than the trivial solution $\psi=0$. In addition, this solution has an effective energy $E_{\rm eff.}\(\psi_c\)$ that is less than or equal to that of any other solution vanishing at infinity.  Any other solution vanishing at infinity which is not both spherically symmetric and monotone has an effective energy that is strictly greater than  $E_{\rm eff.}\(\psi_c\)$.

Note that $E_2$ and $E_{\rm eff.}$ differ  by the topological term $\int d^3x\,\(G+c\)\,\psi^s$, which is constant within a given homotopy class. Therefore, the minima of $E_2$ for a given value of the Skyrme topological charge $Q$ correspond to the minima of $E_{\rm eff.}$ for that same value of $Q$. Consequently, the minima of $E_2$ are monotone and spherically symmetric too. Due to such spherical symmetry we shall use an holomorphic ansatz for the $U$-fields in terms of a radial profile function $f\(r\)$, and a complex scalar field $u$ depending upon the angles, i.e. 
\be
U= W^{\dagger}\, e^{i\,f\, T_3}\,W, \qquad\qquad {\rm with}\qquad \qquad W=\frac{1}{\sqrt{1+\u2}}\(
\begin{array}{cc}
1& i\,u\\
i\, \ub&1
\end{array}\)
\ee 
For such a type of configuration, the self-duality equations are satisfied by an $h$-matrix of the form $h=d\(V\)\,h_D\,d^T\(V\)$, where $d\(V\)$ is the adjoint (triplet) representation of the group element $V=W^{\dagger}\, e^{i\,f\, T_3/2}$, with 
\be
h_D=\(f^{\prime}\(r\)/\lambda\)\,{\rm diag.}\,\left[1\,,\,1\,,\,\frac{4\,\sin^2\(f/2\)}{r^2\,{f^{\prime}}^2}\right]
\ee
and where we have chosen $u=\(x_2+i\,x_1\)/\(r-x_3\)$   (see \cite{us}). In order for the baryonic charge $Q$  to be positive (see \rf{signlambdaq}) we shall choose the self-dual sector where $\lambda=-m_0\,e_0 <0$, and so $f$ is a monotonically decreasing function of $r$. From \rf{psidef}, or equivalently ${\rm det}\,h=4\,\psi^s$, we get that
\br
\psi^s=-\(m_0\,e_0\)^{-3}\,\frac{1}{2\,r^2}\,\frac{d\,}{d\,r}\left[f-\sin f\right]
\lab{psifrel}
\er
which leads, for  the boundary conditions $f\(0\)=2\,\pi\mid Q\mid$ and $f\(\infty\)=0$, to a positive Skyrme topological charge. Note that this does not imply that we are considering the Skyrme product ansatz, since  $f_{\mid Q\mid}\neq \mid Q\mid f_1$, and so $U_{\mid Q\mid}\neq U^{\mid Q\mid}_1$. Note that negative topological charges can be obtained by choosing the other self-dual sector where $\lambda>0$ and $f$ is a monotonically increasing function of $r$. In both cases $\psi$ and the eigenvalues of $h$ are positive. So far, the results do not depend upon the detailed form of the effective potential, as long as it is admissible in the sense of \cite{coleman2}. 

Coleman's false vacuum argument has established that, for each sector of topological charge $Q$,  the minimum of the energy $E_2$, given in \rf{e2}, has to be spherically symmetric and monotone. Therefore,  \rf{psieq} reduces to an ordinary differential equation for a radial function $\psi\(r\)$, i.e.
\be
\mu_0^2\left[\frac{d^2\,\psi}{d\,r^2}+\frac{2}{r}\,\frac{d\,\psi}{d\,r}\right]-\frac{\delta\,V_{\rm eff.}}{\delta\,\psi}=0
\lab{psieqradial}
\ee
and $V_{\rm eff.}$ has to be negative for some interval of values of $\psi$. Since we want $E_2$ to be positive, we shall take $V$ to be positive. Therefore, the integration constant $c$ has to be positive. In order to comply with \rf{effpotcond} and \rf{massscale} we shall take 
\br
V_{\rm eff.}= \beta_2^2\,\psi^2-c\,\psi^s+{\widetilde V};\qquad \qquad2<s<6 
\lab{potminus1}
\er
and
\be
{\widetilde V}\geq 0;\qquad \quad {\widetilde V}\mid_{\psi=0}=0;\qquad \quad\frac{\delta\,{\widetilde V}}{\delta\,\psi}\mid_{\psi=0}=0
;\qquad\quad \frac{\delta^2\,{\widetilde V}}{\delta^2\,\psi}\mid_{\psi=0}=0
\ee
 For sufficiently large $c$ the effective potential $V_{\rm eff.}$ will certainly be negative in some region of $\psi$, and so $V_{\rm eff.}$ will be admissible in the sense of \cite{coleman2}. However, there is a non-negative critical value $c_{\rm crit.}$ such that for $c<c_{\rm crit.}$ the effective potential will be positive everywhere and Coleman's false vacuum solution ceases to exist. 

Note that Coleman's false vacuum solution can exist even for the cases where $c\, \psi^s>{\widetilde V}$, for $\psi\rightarrow \infty$, and so $V_{\rm eff.}$ is unbounded from below. In such cases we have $c_{\rm crit.}=0$, and for the solutions corresponding to $c$ approaching such $c_{\rm crit.}$, the value of $\psi$, at $r=0$, diverges. We shall not consider such situation as we do not want divergent values of the baryonic charge density. Consequently, we shall consider the cases where $c\, \psi^s<{\widetilde V}$, for $\psi\rightarrow \infty$, and so $V_{\rm eff.}$ is bounded from below. Therefore, $V_{\rm eff.}$ has a true global vacuum $\psi_{\rm vac.}$, such that   
\be
\frac{\delta\,V_{\rm eff.}}{\delta\,\psi}\mid_{\psi_{\rm vac.}}=0;\qquad\qquad V_{\rm eff.}\(\psi_{\rm vac.}\)<0
\ee
As $c$ decreases such a vacuum becomes shallow and at a critical value $c_{\rm crit.}>0$, it  becomes degenerated with the false vacuum $\psi=0$, i.e. $\psi_{\rm vac.}\rightarrow \psi_{\rm crit.}$, such that $V_{\rm eff.}\(\psi_{\rm crit.}\)=0$. At this point Coleman's false vacuum solution ceases to exist, and one gets instead a constant solution $\psi=\psi_{\rm crit.}$, with divergent value of the integrated baryonic charge, i.e. $Q\rightarrow \infty$.

As Coleman points out \cite{coleman1,coleman2}, one can analyse \rf{psieqradial} as the problem of a mechanical particle with position being given by the coordinate $\psi$,  as a function of the time $r$, and moving under the influence of the inverted potential $V_{\rm eff.}^{(-)}\equiv -V_{\rm eff.}$, and of a viscous force $-\mu_0^2\,\frac{2}{r}\,\frac{d\,\psi}{d\,r}$. Multiplying \rf{psieqradial} by $\frac{d\,\psi}{d\,r}$, one gets that
\be
\frac{d\,{\cal E}_{\rm particle}}{d\,r}=-\mu_0^2\,\frac{2}{r}\,\(\frac{d\,\psi}{d\,r}\)^2
\lab{eparticleder}
\ee
with
\be
{\cal E}_{\rm particle}=\frac{\mu_0^2}{2}\,\(\frac{d\,\psi}{d\,r}\)^2+V_{\rm eff.}^{(-)}
\lab{eparticle}
\ee
By expanding \rf{psieqradial} around $r=0$ one concludes that $\frac{d\,\psi}{d\,r}\mid_{r=0}=0$. From \rf{eparticleder} one observes that as the particle moves from $r=0$, the energy ${\cal E}_{\rm particle}$ only decreases. Therefore, we have to release the particle at $r=0$, from a position $\psi_0$, such that ${\cal E}_{\rm particle}\(0\)=V_{\rm eff.}^{(-)}\(\psi_0\)>0$, and so with $\psi_0$ inside the interval where $V_{\rm eff.}$ is negative. The Coleman's false vacuum solution will end, as $r\rightarrow \infty$, at $\psi=0$, with zero velocity $\frac{d\psi}{dr}$, and so with zero energy ${\cal E}_{\rm particle}$. So, we have that
\be
{\cal E}_{\rm particle}\(0\)=V_{\rm eff.}^{(-)}\(\psi_0\)=\mu_0^2\,\int_0^{\infty}dr\,\frac{2}{r}\,\(\frac{d\,\psi}{d\,r}\)^2
\lab{eparticlezero}
\ee
Note that as $c$ approaches $c_{\rm crit.}$, from above, one has that $\psi_0\rightarrow \psi_{\rm crit.}$, and so $V_{\rm eff.}^{(-)}\(\psi_0\)\rightarrow 0$. As the integrand on the l.h.s of \rf{eparticlezero} is positive, one observes that it will have to go to zero everywhere as $c \rightarrow c_{\rm crit.}$, from above. But Coleman's false vacuum solutions eventually decay, at large distances, to the false vacuum $\psi=0$, within a length scale $a$, which is independent of $c$, as given in \rf{psiyukawa}. Since $\frac{d\,\psi}{d\,r}$ is non-vanishing in that region, one observes that as  $c \rightarrow c_{\rm crit.}$, that decay will happen for larger and larger values of $r$, in order for the factor $\frac{2}{r}$, in the integrand on the l.h.s of \rf{eparticlezero}, to suppress the derivative of $\psi$. Therefore, the solutions, for $c$ close to $c_{\rm crit.}$, develop a plateau from $r=0$ to $r=R\(c\)$, where $\psi$ is practically constant, and they decay to the false vacuum for  $r>R\(c\)$. Note however that $R\(c\)$ grows as $c$ decreases towards $c_{\rm crit.}$. In addition, the value of $\psi$ on the plateau approaches  $\psi_{\rm crit.}$ as  $c \rightarrow c_{\rm crit.}$. 

From \rf{top0} we have that 
\be
Q\sim  \int d^3x\, \psi^s\sim \psi_0^s \int_0^{R\(c\)} dr\, r^2\sim  \frac{1}{3}\,R\(c\)^3\,\psi_0^s
\ee
One then observes that the baryonic charge $Q$ grows as $c$ decreases towards $c_{\rm crit.}$. In addition, the root mean square (RMS) radius of the solution becomes
\be
\sqrt{\langle r\rangle}=\sqrt{\frac{\int d^3x\, r^2\,\psi^s}{\int d^3x\, \psi^s}}\sim \sqrt{\frac{3}{5}}\,R\(c\); \qquad \rightarrow \qquad \sqrt{\langle r\rangle}\sim Q^{1/3}
\lab{rqonethird}
\ee
Consequently, the solutions for $c$ close to $c_{\rm crit.}$ tend to satisfy a linear relation between the RMS radius and $ Q^{1/3}$, which is the power law one observes experimentally for nuclei above some baryonic charge (mass number). 

One observes that the false vacuum solution presents some robust properties, which are quite independent of the value of the parameters of the model. First, the baryonic charge decreases with the increase of the integration constant $c$. Second, the relation between RMS radius and baryonic charge obeys the relation $\sqrt{\langle r\rangle}\sim Q^{1/3}$. Those two properties are valid as $c$ approaches $c_{\rm crit.}$. However, as we will see below, the range of $c$ where  such properties are valid is large enough to accommodate most of the nuclei. The discrepancies are relevant for baryonic charge below 12.

\section{The binding energy per nucleon}
\label{sec:binding}
\setcounter{equation}{0}

Another important property of our model is that the binding energy per nucleon presents a very robust dependency upon the baryonic charge, for a quite large portion of the parameter space. As we have argued at the end of section \ref{sec:falsevacuum}, the  baryonic charge $Q$ is a monotonic decreasing function of the integration constant $c$, at least for a  region where $c$ approaches the critical value $c_{\rm crit.}$ where the Coleman's false vacuum solution ceases to exist.  Assuming that dependency of $Q$ upon $c$, and for some suitable region of the parameters in $G\(U\)$ (see \rf{e2}), we show that the binding energy per nucleon increases with $Q$, for small values of $Q$, up to a maximum, and then decreases with $Q$, for large values of the baryonic charge. 

As we have seen, Derrick's scaling argument leads  to the relation \rf{derrick0} between the kinetic energy $T$ and the effective potential energy $U_{\rm eff.}$. Combining \rf{derrick0} with  the relation between $V$ and $V_{\rm eff.}$, given in \rf{psieq}, we replace the integral of $V$ in $E_2$, given in \rf{e2}, by  $\int d^3x\,V=\int d^3x\,\(V_{\rm eff.}+c\,\psi^s\)=-\frac{T}{3}+c\,\int d^3x\,\psi^s$, and so
\be
E_2=\frac{2}{3}\,T+\frac{4\,\pi^2}{\mid \lambda\mid^3}\,\,c\,\mid Q\mid+Q_G
\lab{e2t}
\ee
where we have used \rf{topmodulus} and have denoted
\be
Q_G=\int d^3x\,G\(U\)\,\psi^s
\lab{qgdef}
\ee
For a given potential $V$, the solutions of \rf{psieq} are functions of the integration constant $c$. Therefore, differentiating $E_2$, given in \rf{e2}, w.r.t. $c$ and integrating by parts we get
\be
\frac{d\,E_2}{d\,c}=\int d^3x\left[\mu_0^2\,\partial_i\(\partial_i\psi\,\frac{d\,\psi}{d\,c}\)-\mu_0^2\,\partial^2\psi\,\frac{d\,\psi}{d\,c}+\frac{\delta\,V}{\delta\,\psi}\,\frac{d\,\psi}{d\,c}\right]+\frac{d\,Q_G}{d\,c}
\lab{predere2}
\ee
Since $\psi\rightarrow 0$, and $\partial_i\psi\rightarrow 0$ as $r\rightarrow \infty$, in order for the density of baryonic charge to vanish at infinity, we get that the first term on the r.h.s. of \rf{predere2} vanishes. Using, on the second term of \rf{predere2}, the equation \rf{psieq}, and also \rf{topmodulus},  we get that
\be
\frac{d\,E_2}{d\,c}=\frac{4\,\pi^2}{\mid \lambda\mid^3}\,\,c\,\frac{d\mid Q\mid}{d\,c}+\frac{d\,Q_G}{d\,c}
\lab{dere2c}
\ee
The binding energy per nucleon is
\be
E_B= \frac{\mid Q\mid E\(Q=1\)-E\(\mid Q\mid\)}{\mid Q\mid}=\frac{\mid Q\mid E_2\(Q=1\)-E_2\(\mid Q\mid\)}{\mid Q\mid}
\lab{ebdef}
\ee
with $E$ given in \rf{tot}. Therefore, using \rf{e2t} and \rf{dere2c} one gets
\be
\frac{d\,E_B}{d\,c}=\frac{2}{3}\,\frac{T}{\mid Q\mid^2}\,\frac{d\mid Q\mid}{d\,c}-\frac{1}{\mid Q\mid^2}\left[ \mid Q\mid \,\frac{d \, Q_G}{d\,c}-Q_G\,\,\frac{d\mid Q\mid}{d\,c}\right]
\lab{predereb}
\ee
If the relation between $Q$ and $c$ is invertible, i.e. one is a monotonic function of the other, we can write \rf{predereb} as
\be
\frac{d\,E_B}{d\mid Q\mid}=\frac{2}{3}\,\frac{T}{\mid Q\mid^2}\,-\frac{1}{\mid Q\mid^2}\left[ \mid Q\mid \,\frac{d\, Q_G}{d\mid Q\mid}-Q_G\right]
\lab{dereb}
\ee
In order for $E_2$ to be positive we shall take the potential $V$ to be non-negative. However, the Derrick's scaling argument implies \rf{derrick0}, and so $U_{\rm eff.}$ has to be negative, as $T$ is always positive, and so $V_{\rm eff.}$ has to be negative in some region of the $\psi$-axis. The only way to achieve that is to have the integration constant $c$ to be positive, as $\psi$ is positive. From \rf{derrick0} we have that
\be
\frac{T}{3}=\frac{4\,\pi^2}{\mid \lambda\mid^3}\,\,c\mid Q\mid-\int d^3x\,V>0\qquad \rightarrow \qquad 
\frac{4\,\pi^2}{\mid \lambda\mid^3}\,\,c\mid Q\mid>\int d^3x\,V
\ee
Therefore
\be
T=\frac{12\,\pi^2}{\mid \lambda\mid^3}\,\,c\mid Q\mid\,\Omega\qquad {\rm with} \qquad
\Omega =1-\frac{\mid \lambda\mid^3}{4\,\pi^2}\frac{1}{c\mid Q\mid}\,\int d^3x\,V < 1
\ee
We then conclude that
\be
0<\frac{T}{\mid Q\mid^2}<\frac{12\,\pi^2}{\mid \lambda\mid^3}\,\,\frac{c}{\mid Q\mid}
\lab{boundont}
\ee

The type of nuclei we shall consider are such that the number of protons and neutrons do not differ drastically, and so, the charges of such nuclei are approximately proportional to $\mid Q\mid$. Therefore, we shall choose the functional $G\(U\)$ such that $Q_G$ is positive and proportional to $\mid Q\mid^2$, i.e. $Q_G=\chi\, \mid Q\mid^2$ ($\chi>0$), and so $Q_G$ approximates the Coulomb interaction energy for the nuclei. Then, \rf{dereb} becomes 
\be
\frac{d\,E_B}{d\mid Q\mid}=\frac{2}{3}\,\frac{T}{\mid Q\mid^2}\,-\chi
\lab{dereb2}
\ee
In order to obtain \rf{dereb}  we have assumed that $c$ is a monotonic function of $\mid Q\mid$. If we go further and assume that $c$ is a monotonically decreasing function of $\mid Q\mid$, we can get, from \rf{boundont} and \rf{dereb2},  the bulk of the form of the binding energy per nucleon as a function of $\mid Q\mid$. Indeed, with an appropriate value for $\chi$, we get that for sufficiently small values of $\mid Q\mid$ we have that 
$\frac{2}{3}\,\frac{T}{\mid Q\mid^2}\,>\chi$, and so $E_B$ grows with the increase of $\mid Q\mid$, and for sufficiently large values of $\mid Q\mid$, we get that $\frac{2}{3}\,\frac{T}{\mid Q\mid^2}\,<\chi$, and so $E_B$ decreases with the increase of $\mid Q\mid$. However, with such analysis we can not rule out local minima and maxima of $E_B$.

We have shown, at the end of section \ref{sec:falsevacuum}, that the  baryonic charge $Q$ is a monotonic decreasing function of the integration constant $c$, at least for a  region where $c$ approaches the critical value $c_{\rm crit.}$. As we will see below, that region can be large enough to accommodate all nuclei. Therefore the dynamics of our model leads in a quite robust way for a binding energy per nucleon that reproduces the bulk of the features of the experimental data. The choice of the parameters of $V$  and $G\(U\)$ can be used to match the scales and so to fine tune that experimental data, as we now explain.

\section{Choice of the potential}
\label{sec:potential}
\setcounter{equation}{0}

We shall consider the following  potential, admissible in the sense of  \cite{coleman1,coleman2},
\br
V_{\rm eff.}= \beta_2^2\,\psi^2-c\,\psi^s+\beta_{\kappa}^2\,\psi^{\kappa};\quad 6>s>2;\;\;\kappa>s \lab{pot0}
\lab{especificeffpot}
\er
with the ratio $\mu_0/\beta_2$  determined from \rf{massscale}, as
\be
\frac{\mu_0}{\beta_2}=\sqrt{2}\,s\,a
\lab{mu0beta2a}
\ee
 In order to better analyze the role of the parameters of the model, let us  define the dimensionless quantities 
\be 
\zeta=\frac{\beta_2}{\mu_0}\,r;\qquad\qquad\hpsi=\(\frac{\beta_{\kappa}^2}{\beta_2^2}\)^{\frac{1}{\(\kappa-2\)}}\,\psi ;\qquad\qquad \gamma=\frac{c}{\beta_2^2} \,\(\frac{\beta_2^2}{\beta_{\kappa}^2}\)^{\frac{s-2}{\kappa-2}} \lab{dimensionless}
\ee
According to Coleman's false vacuum argument the solution of \rf{psieq} with smallest energy has to be radial, and so using \rf{dimensionless}, one gets that \rf{psieq} becomes 
\br
\frac{d^2\,\hpsi}{d\,\zeta^2}+\frac{2}{\zeta}\,\frac{d\,\hpsi}{d\,\zeta}- \frac{\delta \widehat{V}_{{\rm eff.}}}{\delta \hpsi}=0;\qquad\qquad \widehat{V}_{{\rm eff.}}=\widehat{V}-\gamma\,\hpsi^{s}=\hpsi^2-\gamma\,\hpsi^{s}+\hpsi^{\kappa}
\lab{hpsieq}
\er
In addition,  the topological charge becomes
\br
\mid Q\mid=\(\frac{\mu_0}{\beta_2}\)^3\,\vartheta\; I\(\gamma\,,\,s\,,\,\kappa\)
\lab{chargefinal}
\er
with 
\be
\vartheta=\(m_0\,e_0\)^3\,\(\frac{\beta_2^2}{\beta_{\kappa}^2}\)^{\frac{s}{\(\kappa-2\)}}\quad \qquad\qquad  I=\frac{1}{\pi}\int_0^{\infty}d\zeta\,\zeta^2\,\hpsi^s
\lab{varthetaIdef}
\ee 
Therefore, the density of baryonic charge becomes
\be
\rho=\frac{\vartheta}{4\,\pi^2}\,\hpsi^s\; \qquad\qquad {\rm with}\qquad\qquad \mid Q\mid=\int d^3x\, \rho
\lab{densitypsihat}
\ee
The root-square-mean radius of the baryonic charge is
\be
\sqrt{\langle r^2\rangle}\equiv\sqrt{\frac{\int d^3x\, r^2\,\psi^s}{\int d^3x\, \psi^s}}=\frac{\mu_0}{\beta_2}\,\Lambda\(\gamma\,,\,s\,,\,\kappa\)
\lab{radius}
 \ee 
 with 
 \be
 \Lambda \equiv \sqrt{\frac{J\(\gamma\,,\,s\,,\,\kappa\)}{I\(\gamma\,,\,s\,,\,\kappa\)}};\qquad \qquad 
 J=\frac{1}{\pi}\int_0^{\infty}d\zeta\,\zeta^4\,\hpsi^s
 \lab{LambdaJdef}
 \ee
We look for solutions of \rf{hpsieq} satisfying $\hpsi'\(0\)=0$ and $\hpsi\(\infty\)=\hpsi'\(\infty\)=0$. For given values of $\kappa$ and $s$, we can vary the parameter $\gamma$, which has been traded by the arbitrary integration constant $c$, and obtain solutions of any baryonic charge.   Note that the ratio $\mu_0/\beta_2$ gives only the scale of the rms radii \rf{radius} and the shape is given only by $\Lambda\(s,\,\kappa,\,\gamma\)$. 

Note that $\hpsi=0$ is a local minimum of the effective potential \rf{especificeffpot}, i.e. the false vacuum. The global minimum, or the true vacuum, occurs for some positive value of $\hpsi$. However, as $\gamma$ decreases there will be a critical value $\gamma_{\rm crit.}$ such that the true vacuum becomes degenerated with the false vacuum, and Coleman's false vacuum solution ceases to exist, as we discussed in section \ref{sec:falsevacuum}. That happens when 
\be
V_{\rm eff.}\mid_{\gamma_{\rm crit.}\,,\,\hpsi_{\rm crit.}}=0;\qquad\qquad \frac{\delta \, V_{\rm eff.}}{\delta\, \hpsi}\mid_{\gamma_{\rm crit.}\,,\,\hpsi_{\rm crit.}}=0
\ee
which implies
\be 
\gamma_{\rm crit.}= \(\frac{\kappa-2}{\kappa-s}\)\,\(\frac{\kappa-s}{s-2}\)^\frac{s-2}{\kappa-2};\qquad\qquad \hpsi_{\rm crit.}= \(\frac{s-2}{\kappa-s}\)^\frac{1}{\kappa-2} 
\lab{critical}
\ee

\subsection{The bulk properties of the model}

Given the values of $s$ and $\kappa$ we can solve the equation \rf{hpsieq} to find Coleman's false vacuum solution $\hpsi$, for several values of the re-scaled integration constant $\gamma$. Then we can evaluate the quantities $I$ and $\Lambda$, given in \rf{varthetaIdef}  and \rf{LambdaJdef} respectively, which are the re-scaled baryonic charge $\mid Q\mid$ and root-square-mean radius $\sqrt{\langle r^2\rangle}$, respectively. 

\begin{figure}[H]
\begin{center}
		\includegraphics[scale=0.5]{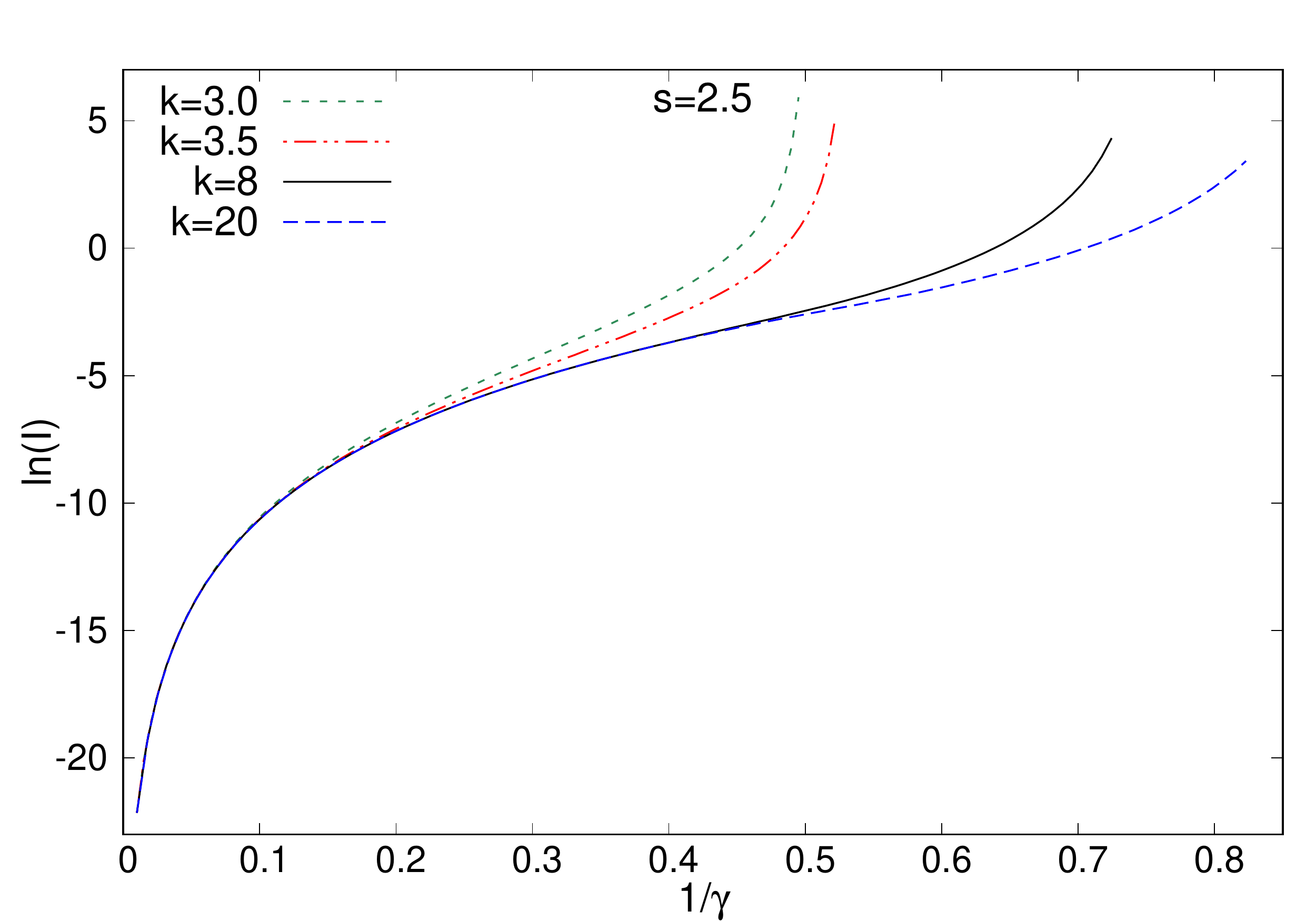}
			\caption{The logarithm of the quantity $I$, defined in \rf{varthetaIdef}, against $1/\gamma$,  for $s=2.5$ and $\kappa=3.0,\,3.5,\,8,\,20$.}
		\label{fig:s=5_2a}
\end{center} 
\end{figure}

\begin{figure}[H]
\begin{center}
		\includegraphics[scale=0.5]{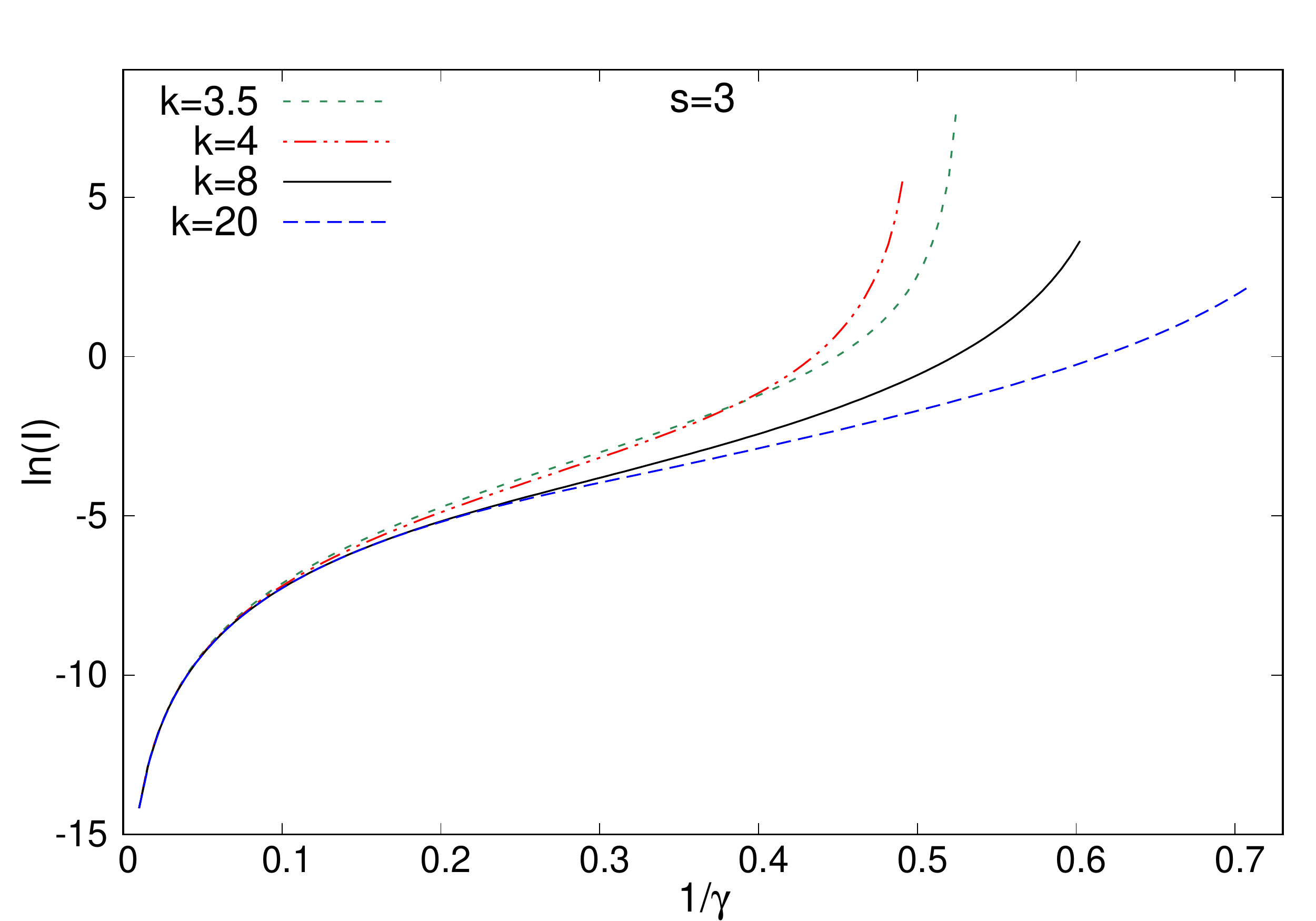}
		\caption{The logarithm of the quantity $I$, defined in \rf{varthetaIdef}, against $1/\gamma$,  for $s=3$ and $\kappa=3.5,\,4,\,8,\,20$.}
		\label{fig:s=3a}
\end{center} 
\end{figure}

\begin{figure}[H]
\begin{center}
		\includegraphics[scale=0.5]{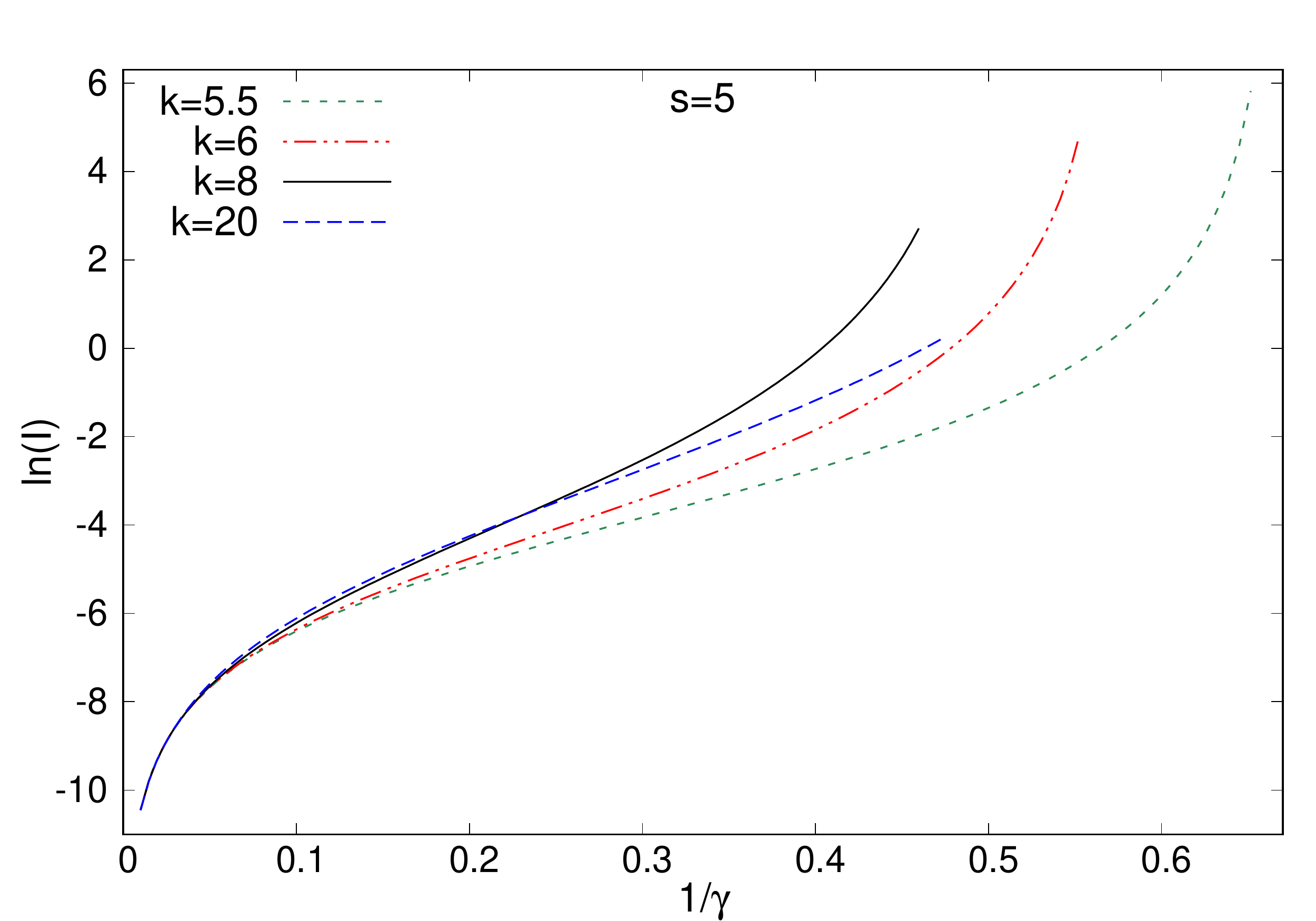}
		\caption{The logarithm of the quantity $I$, defined in \rf{varthetaIdef}, against $1/\gamma$,  for $s=5$ and $\kappa=5.5,\,6,\,8,\,20$.}
		\label{fig:s=5a}
\end{center} 
\end{figure}

In Figures \ref{fig:s=5_2a}, \ref{fig:s=3a} and \ref{fig:s=5a}, we plot $\ln \(I\)$ against $1/\gamma$, for several values of $s$ and $\kappa$. One can clearly observes that the quantity $I$, and so the baryonic charge $\mid Q\mid$, is a monotonically decreasing function of the re-scaled integration constant $\gamma$. In addition, the quantity $I$ grows very fast as the integration constant $\gamma$ decreases to its critical value $\gamma_{\rm crit.}$, given in \rf{critical}. For very large values of $\gamma$, the quantity $I$ becomes very small and insensitive to the value of $\kappa$. Indeed, that can be understood from the equation \rf{hpsieq} for $\hpsi$. Re-scale $\hpsi$ as 
\be
{\bar \psi}= \gamma^{1/(s-2)}\,\hpsi
\lab{barpsidef}
\ee
Then, \rf{hpsieq} becomes
\be
\frac{d^2\,{\bar \psi}}{d\,\zeta^2}+\frac{2}{\zeta}\,\frac{d\,{\bar \psi}}{d\,\zeta}=2\,{\bar \psi}-s\,{\bar \psi}^{s-1}+\frac{\kappa}{\gamma^{(\kappa-2)/(s-2)}}\,{\bar \psi}^{\kappa-1}
\lab{barpsieq}
\ee
Since we are assuming $\kappa>s$ (see \rf{pot0}), we observe that in the limit $\gamma\rightarrow \infty$, the term ${\bar \psi}^{\kappa-1}$ decouples from the equation, and the solutions of \rf{barpsieq} depend upon $s$ only. As we are assuming $s>2$ (see \rf{pot0}) we get from \rf{barpsidef} that for ${\bar \psi}$ to be finite one needs $\hpsi\rightarrow 0$, as $\gamma\rightarrow \infty$. So, indeed $I\rightarrow 0$, as $\gamma\rightarrow \infty$, as we observe from Figures \ref{fig:s=5_2a}, \ref{fig:s=3a} and \ref{fig:s=5a}. 

The root-square-mean radius of the baryonic charge, given in \rf{radius}, becomes
\be
\sqrt{\langle r^2\rangle}\mid_{\gamma\rightarrow\infty}=\sqrt{\frac{\int d^3x\, r^2\,\psi^s}{\int d^3x\, \psi^s}}\mid_{\gamma\rightarrow\infty}=\frac{\mu_0}{\beta_2}\,{\bar \Lambda}\(s\)\equiv r_{\rm min.}\(s\)
\lab{radiusbiggamma}
 \ee 
 with
  \be
{\bar  \Lambda} \equiv \sqrt{\frac{{\bar J}\(s\)}{{\bar I}\(s\)}};\qquad 
 {\bar I}\(s\)=\frac{1}{\pi}\int_0^{\infty}d\zeta\,\zeta^2\,{\bar \psi}^s;\qquad 
 {\bar J}\(s\)=\frac{1}{\pi}\int_0^{\infty}d\zeta\,\zeta^4\,{\bar \psi}^s;\qquad \gamma\rightarrow \infty
 \lab{LambdaJbardef}
 \ee

 Note that using the relations \rf{chargefinal}, \rf{radius} and \rf{radiusbiggamma}  we can express the rms radius as 
\be
\sqrt{\langle r^2\rangle}= r_{\rm min.}\(s\)+\,\vartheta^{-1/3}\,\,\Omega\(\gamma\,,\,s\,,\,\kappa\) \mid Q\mid^{1/3}
\lab{radiuschargerel}
 \ee 
 with
 \be
 \Omega\(\gamma\,,\,s\,,\,\kappa\)\equiv \sqrt{\frac{J\(\gamma\,,\,s\,,\,\kappa\)}{I^{5/3}\(\gamma\,,\,s\,,\,\kappa\)}}-\frac{{\bar \Lambda}\(s\)}{I^{1/3}\(\gamma\,,\,s\,,\,\kappa\)}
 \ee
 Note that
 \be
 \Omega\(\gamma\,,\,s\,,\,\kappa\)\rightarrow 0; \qquad\qquad {\rm as}\qquad \gamma\rightarrow\infty
 \ee
For $\gamma=\gamma_{\rm crit.}$ we get that $\hpsi$ becomes constant and equal to $\hpsi_{\rm crit.}$ given by \rf{critical}. Therefore, 
\be
J\rightarrow \frac{ \hpsi_{\rm crit.}^{\,s}}{\pi}\,\int_0^R d\zeta\, \zeta^4;\qquad
I\rightarrow  \frac{\hpsi_{\rm crit.}^{\,s}}{\pi}\,\int_0^R d\zeta\, \zeta^2;\qquad {\rm for}\qquad R\rightarrow\infty
\ee
and so
\be
 \Omega\(\gamma_{\rm crit.}\,,\,s\,,\,\kappa\)=\sqrt{\frac{3^{5/3}}{5}}\, \pi^{1/3}\,\(\frac{\kappa-s}{s-2}\)^\frac{s}{3\(\kappa-2\)} 
 \ee

Note that such  results are valid for any choice of the coupling constants, and so the fact that the baryonic charge $\mid Q\mid$ is a monotonically decreasing function of $\gamma$, is a quite robust property of our model. Such property was the basic assumption we have made in section \ref{sec:binding} to show that the binding energy per nucleon has the desired dependency upon the baryonic charge.  The choice of the values of the coupling constants will be necessary solely to fix the scales, and not the bulk behavior of the binding energy per nucleon.  

\begin{figure}[H]
\begin{center}
		\includegraphics[scale=0.5]{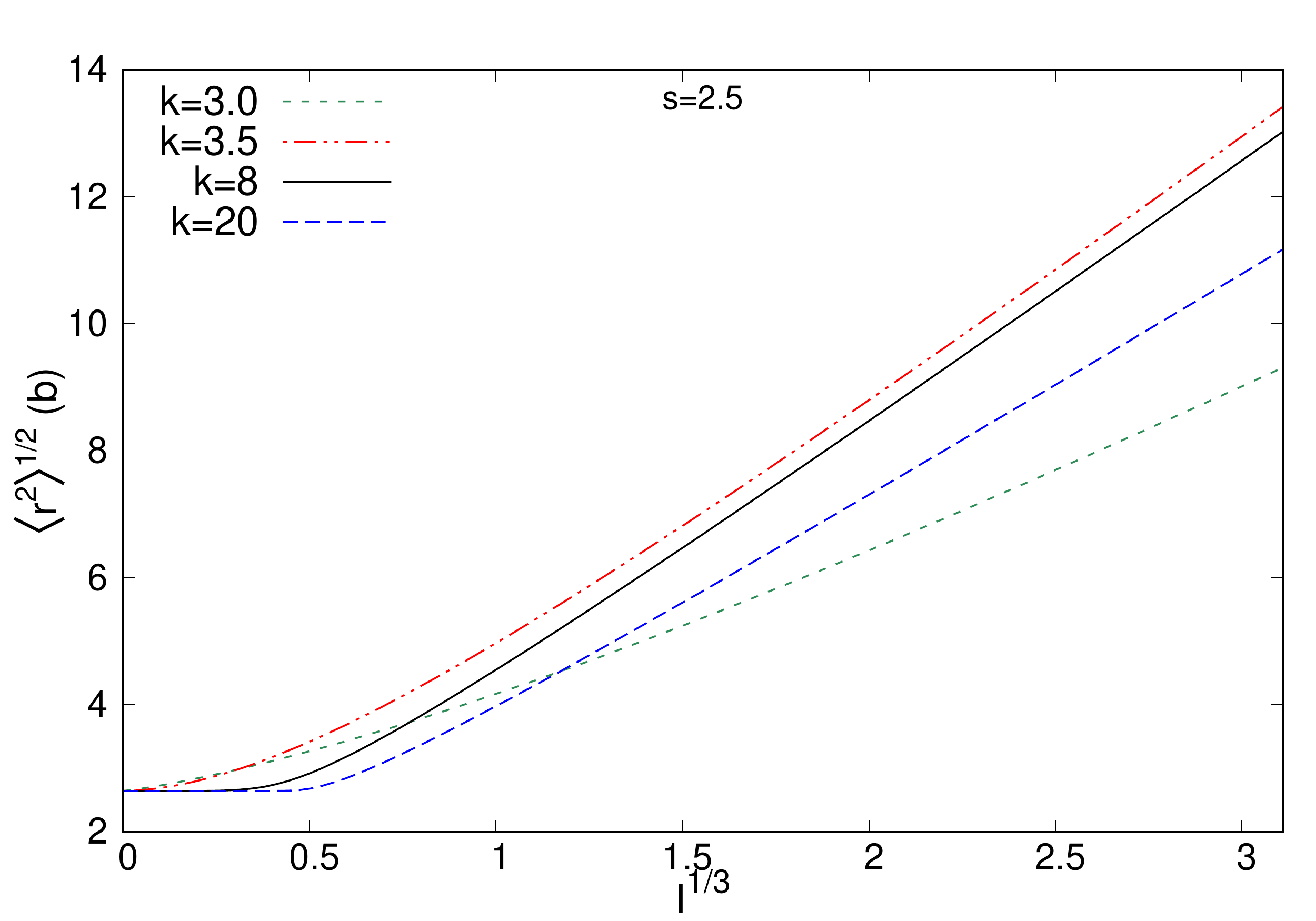}
			\caption{The rms radii \rf{radius}, in units of $b\equiv \frac{a}{0.524}=\frac{\mu_0}{\beta_2}\,\frac{1}{\sqrt{2}\,s\, 0.524}$, with $a$ given in \rf{massscale}, against $I^{1/3}$, for $s=2.5$ and $\kappa=3.0,\,4.5,\,8,\,20$.}
		\label{fig:s=2.5}
\end{center} 
\end{figure}

\begin{figure}[H]
\begin{center}
		\includegraphics[scale=0.5]{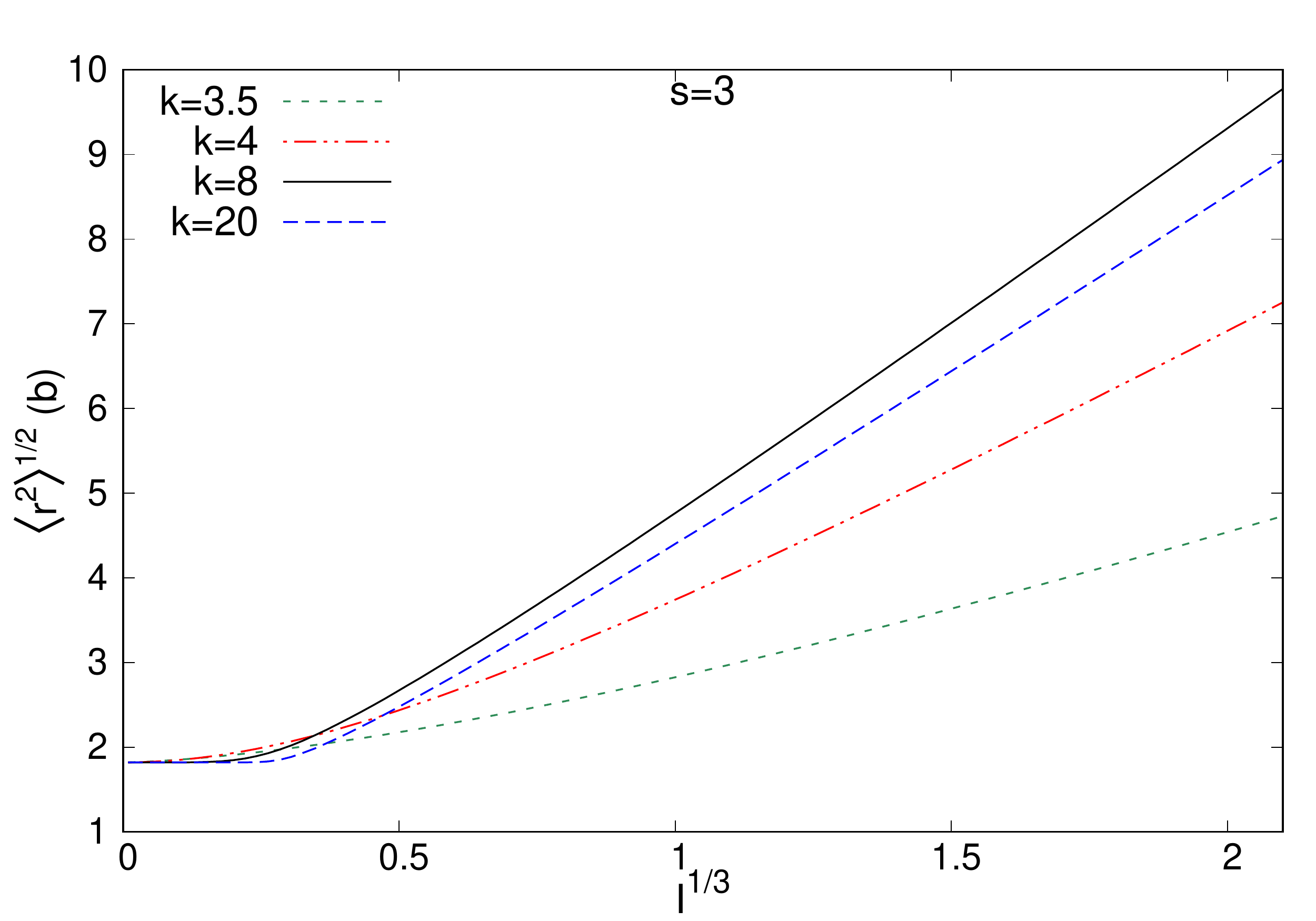}
			\caption{The rms radii \rf{radius}, in units of $b\equiv \frac{a}{0.524}=\frac{\mu_0}{\beta_2}\,\frac{1}{\sqrt{2}\,s\, 0.524}$, with $a$ given in \rf{massscale}, against  $I^{1/3}$, for $s=3$ and $\kappa=3.5,\,4,\,8,\,20$.}
		\label{fig:s=3}
\end{center} 
\end{figure}

\begin{figure}[H]
\begin{center}
		\includegraphics[scale=0.5]{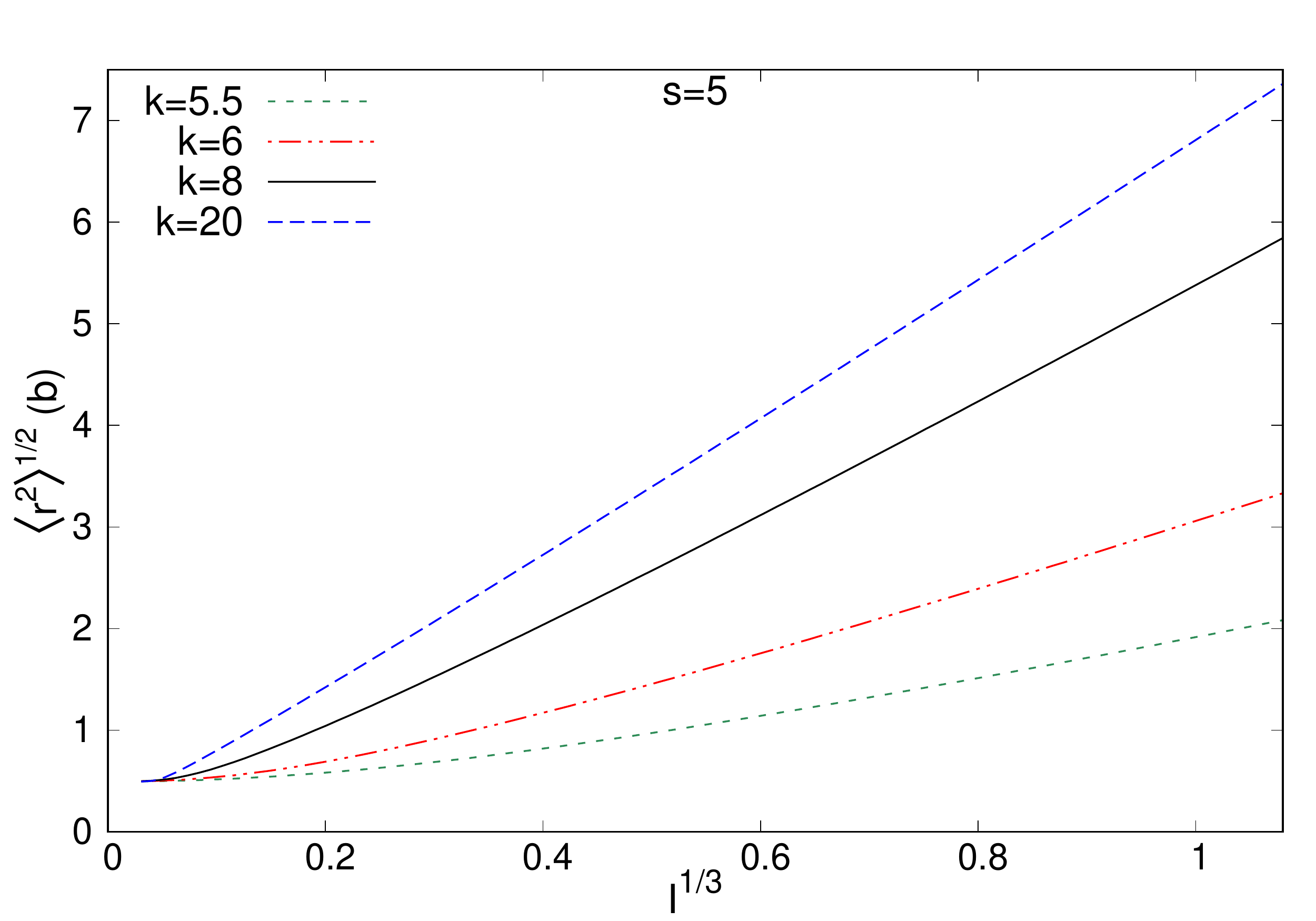}
			\caption{The rms radii \rf{radius}, in units of $b\equiv \frac{a}{0.524}=\frac{\mu_0}{\beta_2}\,\frac{1}{\sqrt{2}\,s\, 0.524}$, with $a$ given in \rf{massscale},  against  $I^{1/3}$, for $s=5$ and $\kappa=5.5,\,6,\,8,\,20$.}
		\label{fig:s=5}
\end{center} 
\end{figure}

In Figures \ref{fig:s=2.5}, \ref{fig:s=3} and \ref{fig:s=5}, we plot the root-square-mean radius, given in \rf{radius}, in units of  $b\equiv \frac{a}{0.524}=\frac{\mu_0}{\beta_2}\,\frac{1}{\sqrt{2}\,s\, 0.524}$, or equivalently $\sqrt{2}\,s\, 0.524\,\Lambda$, against $I^{1/3}$, with $a$ given in \rf{massscale} and $I$ given in \rf{varthetaIdef}, for several values of $s$ and $\kappa$. Note that the rms radii becomes a linear function of $I^{1/3}$, and so of $\mid Q\mid^{1/3}$, very quickly, and as $I$ increases, such linearity reinforces itself. That is in complete agreement with the arguments that we presented in section \ref{sec:falsevacuum}, leading to the relation \rf{rqonethird}. Indeed, as the integration constant $\gamma$ approaches its critical value $\gamma_{\rm crit.}$ the rms radii scales as $\mid Q\mid^{1/3}$. In order to see that  such a thing indeed happens close to $\gamma_{\rm crit.}$, one should observe that in the three plots in Figures \ref{fig:s=2.5}, \ref{fig:s=3} and \ref{fig:s=5}, the linear relation becomes reasonably valid for $I^{1/3}>0.5$, which corresponds to $\ln\(I\)>-2.08$. From Figures \ref{fig:s=5_2a}, \ref{fig:s=3a} and \ref{fig:s=5a}, one observes that it corresponds to $\gamma<3.3$, and so reasonably inside the region right above the critical value $\gamma_{\rm crit.}$, given in \rf{critical}.

Note from Figures \ref{fig:s=2.5}, \ref{fig:s=3} and \ref{fig:s=5} that as $I\rightarrow 0$, and so $\gamma\rightarrow \infty$, the radius goes to a non-zero constant independent of $\kappa$. That is the value of the radius we give in \rf{radiusbiggamma}. For  $s=2.5,\,3$ and $5$ they are respectively $2.6437$, $1.8195$ and $0.4949$, in units of $b\equiv \frac{a}{0.524}=\frac{\mu_0}{\beta_2}\,\frac{1}{\sqrt{2}\,s\, 0.524}$,  and they correspond to the numerical limit $I\rightarrow 0$ in the Figures \ref{fig:s=2.5}-\ref{fig:s=5}.

Again, all such results are valid for any choice of the coupling constants, as we have not fixed the values of any of them so far. Therefore, the fact that the root-square-mean radius depends linearly upon $\mid Q\mid^{1/3}$ is a quite robust property of our model. The choice of $s$ and $\kappa$ will be used to fix the slope of that linear relation, and so makes it compatible with the experimental data.


 \subsection{The choice of the functional $G$ and the binding energy per nucleon}
 
Note that if we take  $G$, introduced in \rf{e2}, as a function of $g\equiv\(f-\sin f\)$, we have  that it is a functional of ${\rm Tr}\, U$, and so the symmetry $SU(2)_L\otimes SU(2)_R$ of \rf{e1} is broken to the diagonal $SU(2)$ subgroup. In addition, using \rf{psifrel}  we get that 
\be
\int d^3x\, G\,\psi^s= - \frac{2\,\pi}{\(m_0\,e_0\)^{3}}\,\left[F\(g\(\infty\)\)-F\(g\(0\)\) \right];\qquad {\rm with}\qquad \frac{d\,F}{d\,g}=G
\ee
 From the boundary conditions for $f$ we observe  that such a quantity is a functional of $\mid Q\mid$. As we have shown in Section \ref{sec:binding},   if $Q$ is a monotonic function of $c$, which is supported numerically for  plenty of distinct admissible potentials $V_{{\rm eff.}}$, then the binding energy per nucleon \rf{ebdef} associated with the static energy \rf{tot} with $G=0$ is monotonically increasing with $Q$ and saturates at $Q\rightarrow \infty$. As we have argued on the paragraph before \rf{dereb2}, taking $G$ proportional to $\mid Q \mid^2$, it accounts for the Coulomb repulsive interaction of the protons, for the type of nuclei we are considering.  So, we choose $G$ as  $G=-\beta_G^2\,\(f-\sin f\)/\(2\,\pi\) $, and from \rf{e2} we obtain 
 \be 
 E_2=\sigma_2\,\left[\vartheta\,\(\frac{\mu_0}{\beta_2}\)^3\, {\cal E}+ \frac{\sigma_G}{2}\mid Q\mid^2
\right]
\ee
where  
\be
\sigma_2=\frac{4\,\pi^2\,\beta_2^2}{m_0^3\,e_0^3}\,\(\frac{\beta_{\kappa}^2}{\beta_2^2}\)^{\frac{s-2}{\kappa-2}};\qquad   
\sigma_G=\frac{\beta_G^2}{\beta_2^2}\,\(\frac{\beta_2^2}{\beta_{\kappa}^2}\)^{\frac{s-2}{\kappa-2}}
\ee
 and 
 \be
 {\cal E}=\frac{1}{\pi}\int_{0}^{\infty}d\zeta\,\zeta^2\left[\frac{\(\hpsi^{\prime}\)^2}{2}+\hpsi^2+\hpsi^{\kappa}\right]
 \lab{energycal}
 \ee
  So, the binding energy per nucleon \rf{ebdef} becomes 
\be
E_B=\sigma_2\,\left[\vartheta\(\frac{\mu_0}{\beta_2}\)^3\( {\cal E}_{Q=1}-\frac{{\cal E}}{\mid Q\mid}\)+\frac{\sigma_G}{2}\(1-\mid Q\mid\)
\right]
\lab{binding}
\ee

 \section{The data analysis}
\label{sec:data}
\setcounter{equation}{0}
 
 We consider a list of $N_c=265$ nuclei, containing all the  stable nuclei up to $^{208}$Pb, and above that, nuclei up to $^{240}$Pu with a half-life greater than $10^3$ years, according to \cite{nubase2016}. We have a gap in the mass number $A$ from $A=210$ to $A=225$, since there are no nuclei matching those criteria of stability and half-life.  We shall use the experimental values of charge radii as given by \cite{angeli}, and the values of binding energies per nucleon as given in \cite{ame2016}. The complete list  of the $265$ nuclei with their experimental radii and binding energies are given in the table of Appendix \ref{sec:longtable}.
  
As we have seen in \rf{massscale}, our model predicts that the density of nuclear matter falls to zero, at large distances, with a rate independent of the mass number. That is in good agreement with experimental facts. Indeed, a measure of that is the so-called skin thickness parameter $t$, which is defined as the distance over which the charge density falls from $90\%$ of its central value to $10\%$. The value of $t$ is practically independent of the mass number, and it is approximately given by $2.3$ Fermi. Using the two-parameter Fermi model for nuclear matter, i.e. $\rho\(r\)=\rho_0/\(1+e^{\(r-c\)/a}\)$, one gets that $t=a\,4\,\ln 3$, and so $a$ is approximately equal to $0.524$ Fermi. Therefore, from \rf{psiyukawa}, \rf{massscale} and \rf{mu0beta2a},  we shall set
 \be 
a=0.524\; fm;\qquad\qquad\qquad \frac{\mu_0}{\beta_2}=s\,\sqrt{2}\,  0.524\, fm
\lab{fixingbeta0mu0}
 \ee
 Then the root-square-mean radius of the baryonic charge, given in  \rf{radius}, has its scale fixed, and we are left to choose what value of $\gamma$ corresponds to a given nuclei. We shall choose  $^{56}$Fe as the reference nucleus, and so we have $Q\({^{56}{\rm Fe}}\)=56$ and  $\sqrt{\langle r^2\rangle}\({^{56}{\rm Fe} }\)=3.7377$ fm, according to \cite{angeli}. We then find the value of $\gamma$ corresponding to the reference nucleus $^{56}$Fe by solving the following relation, obtained from \rf{radius}, 
 \be
 3.7377\, fm=s\,\sqrt{2}\,0.524\,fm\,\sqrt{\frac{J\(\gamma\({^{56}{\rm Fe}}\)\,,\,s\,,\,\kappa\)}{I\(\gamma\({^{56}{\rm Fe}}\)\,,\,s\,,\,\kappa\)}}
 \lab{radiusfe56}
 \ee
 Note that the function $\sqrt{J/I}$ is a monotonic function of $\gamma$, and so \rf{radiusfe56} is a well-defined condition for finding $\gamma\({^{56}{\rm Fe}}\)$. 
 
 The value of $\vartheta$ is then fixed by the relation \rf{chargefinal}, through
 \be  
56=s^3\,\(\sqrt{2}\,0.524\,fm\)^3\,\vartheta\; I\(\gamma\({^{56}{\rm Fe}}\)\,,\,s\,,\,\kappa\)
  \lab{chargefe56}
  \ee
Once that is done, we  find the discrete sequence of values of $\gamma$ that lead to  integer values of the baryonic charges $Q$, given by \rf{chargefinal}. With that sequence we can find the rms radii $R\equiv\sqrt{\langle r^2\rangle}$ for the nuclei corresponding to those values of charge. We have observed that the slope of the curve $R\times  Q^{1/3}$ depends stronger on  $s$ than on $\kappa$, and the best fit to the experimental data corresponds to $s=3$. We present such result in the Figure \ref{fig:radii}, and the numerical values of radii are given in the table of Appendix \ref{sec:longtable}. 
  
\begin{figure}[H]
\begin{center}
		\includegraphics[scale=0.6]{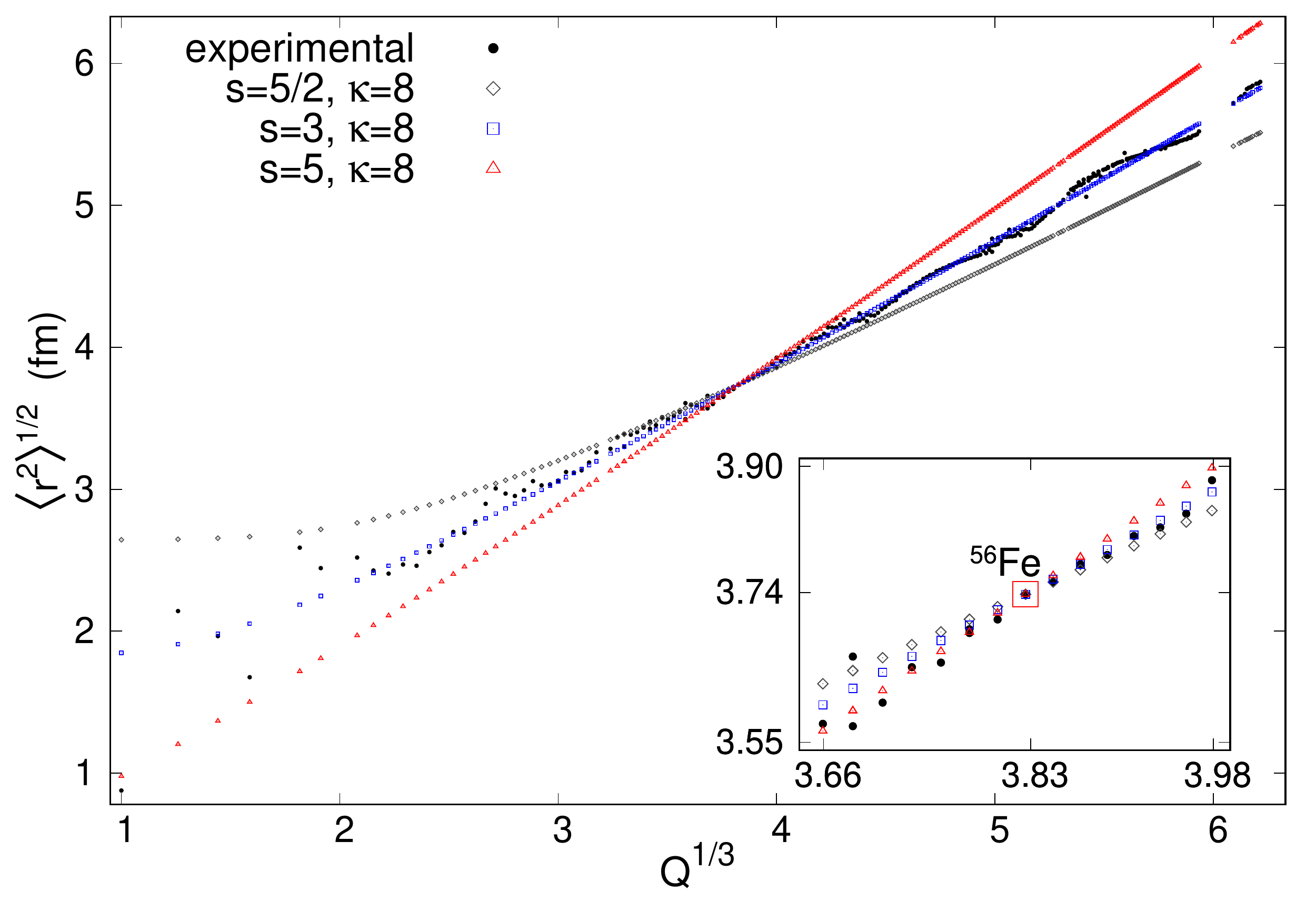}
		\caption{The rms radii $\sqrt{\langle r^2\rangle}$ as a function of $Q^{1/3}$, with $\kappa=8$ and $s=5/2,\,3,\,5$.}
		\label{fig:radii}
\end{center} 
\end{figure}

The experimental and the numerical data are compared using the RMSD of the rms radii $R\equiv \sqrt{\langle r^2\rangle}$ and the binding energy per nucleon $E_B$, defined respectively by
\be \Delta\,R=\sum_{{\rm list}}\sqrt{\frac{\(R^{{\rm num.}}-R^{{\rm exp.}}\)^2}{N_c}};\qquad\qquad \Delta\,E_B=\sum_{{\rm list}}\sqrt{\frac{\(E_B^{{\rm num.}}-E_B^{{\rm exp.}}\)^2}{N_c}} \lab{rmsd}
\ee
where the sum in \rf{rmsd} is over the entire list of $N_c=265$ nuclei defined above. The numerical values $R^{{\rm num.}}$  and $E^{{\rm num.}}$ are obtained respectively  from \rf{radius} and \rf{binding} and the experimental values $R^{{\rm exp.}}$ and $E_B^{{\rm exp.}}$ are given respectively in \cite{angeli} and \cite{ame2016}. We also define $\Delta\,R^{A \geq 12}$ and $\Delta\,E_B^{A \geq 12}$ by restricting the sum of \rf{rmsd} to $A \geq 12$ and taking $N_c^{A \geq 12}=256$, where the lightest nucleus is $^{12}$C.  We have found that,  for $s=3$, $\Delta\,R$ varies from $0.1287\, fm$ for $\kappa=4$, to $0.0837\, fm$ for $\kappa=20$, with a very shallow minimum at $0.0820\, fm$ for $\kappa=8$.   

\begin{figure}[H]
\begin{center}
		\includegraphics[scale=0.6]{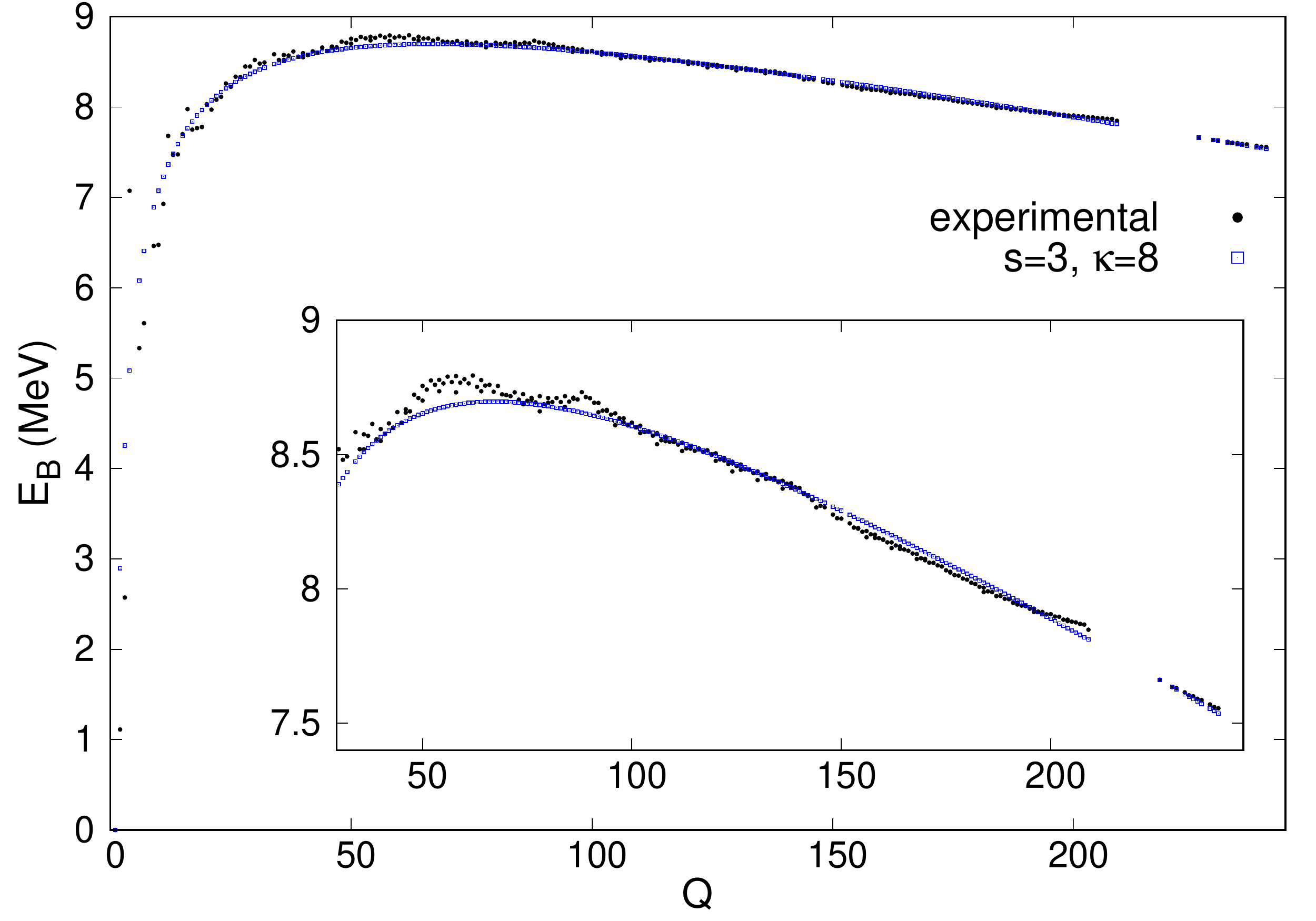}
		\caption{The binding energy per nucleon $E_B$, as a function of $Q$, for $\kappa=8$ and $s=3$, and the best fit corresponding to $\sigma_2=2.16499\,MeV,\, \sigma_G=0.00989002$ where the RMSD is $\Delta E_B =0.216247\, MeV$. }
		\label{fig:binding}
\end{center} 
\end{figure}

We restrict to the case $s=3$, $\kappa=8$, since it gives a very good fit to the radii experimental data. As the values of $\vartheta$ and $\mu_0/\beta_2$ have been fixed already, that also fixes the discrete sequence of values of $\gamma$, and so the value of ${\cal E}_{Q=1}$ (see \rf{energycal} and \rf{binding}). Therefore, we look for the values of $\sigma_2$ and $\sigma_G$,  that give the best fit to the experimental values of the binding energies per nucleon as given in \cite{ame2016}, and find that
\be
\sigma_2=2.16499\,MeV;\qquad \qquad \sigma_G=0.00989002; \qquad\qquad \Delta E_B =0.216247\, MeV
\ee
 The experimental and theoretical results are shown in  Figure \ref{fig:binding}, and they agree with quite good accuracy ($1\%$ or less for $Q>20$).  Once the binding energy is given we can determine the ratio $m_0/e_0$ from  the proton mass, i.e. 
\be
m_p\,c^2=938.272081\;MeV=E\(Q=1\)=48\,\pi^2\,\(m_0/e_0\)+E_2\(Q=1\)
\ee
and so 
\be
m_0/e_0=1.94999\;MeV
\ee
Then, all parameters of our model are fixed except for the ratio $\beta_{\kappa}/\beta_2$. So, the original parameters of the model become 
\br
m_0^2/\alpha&=&4.40047 \;MeV/fm; \qquad \qquad\;\; e_0^2/\alpha=1.15727\;MeV^{-1}\,fm^{-1}
\nonumber\\
\mu_0^2/\alpha^2&=&3.11483\;MeV/fm;\qquad\qquad \beta_2^2/\alpha^2=0.63023\;MeV/fm^3
\\
\beta_G^2/\alpha^3&=&0.00623\;MeV/fm^3;\qquad \qquad{\rm with}\qquad \alpha= \(\beta_{\kappa}/\beta_2\)^{\frac{1}{3}}
\nonumber
\er 

\begin{figure}[H]
\begin{center}
		\includegraphics[scale=0.6]{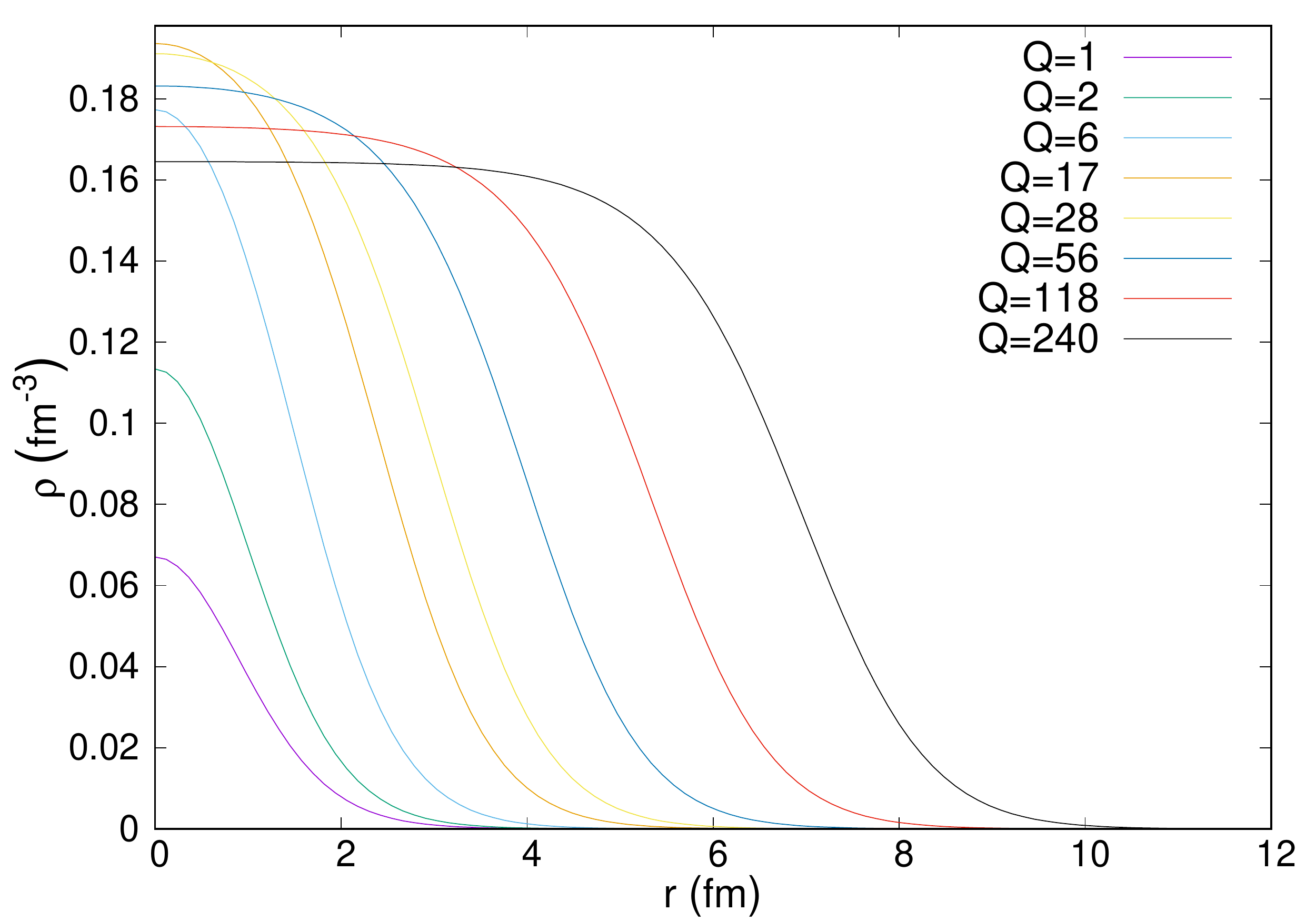}
			\caption{The topological charge density $\rho = \frac{\vartheta}{4\,\pi^2}\hpsi^s\(\zeta=r/\(s\,\sqrt{2}\,0.524\,fm\)\)$ corresponding to the numerical solutions shown in Figure \ref{fig:radii}, for the parameters $\kappa=8$ and $s=3$. Note that the value of $\rho\(r=0\)$ increases up to $Q = 17$ and then decreases slowly to about $0.1645\; fm^{-3}$ for $Q=240$.}
		\label{fig:density}
\end{center} 
\end{figure}

In Figure \ref{fig:density} we plot the density of the baryonic charge as given by \rf{densitypsihat}, in units of $fm^{-3}$, as a function of the radial distance $r$, in units of $fm$, for several values of the baryonic charge. Note that the value of the density at the origin increases up to $Q=17$, and then it decreases slowly to $0.1645\; fm^{-3}$, for  $Q=240$, and that is very close to the experimental values for heavy nuclei given in many text books.  The PREX Collaboration has recently obtained that the interior baryon density of the nucleus $^{208}$Pb is $0.1480\pm  0.0038\,fm^{-3}$ \cite{prex}.

\begin{figure}[H]
\begin{center}
		\includegraphics[scale=0.6]{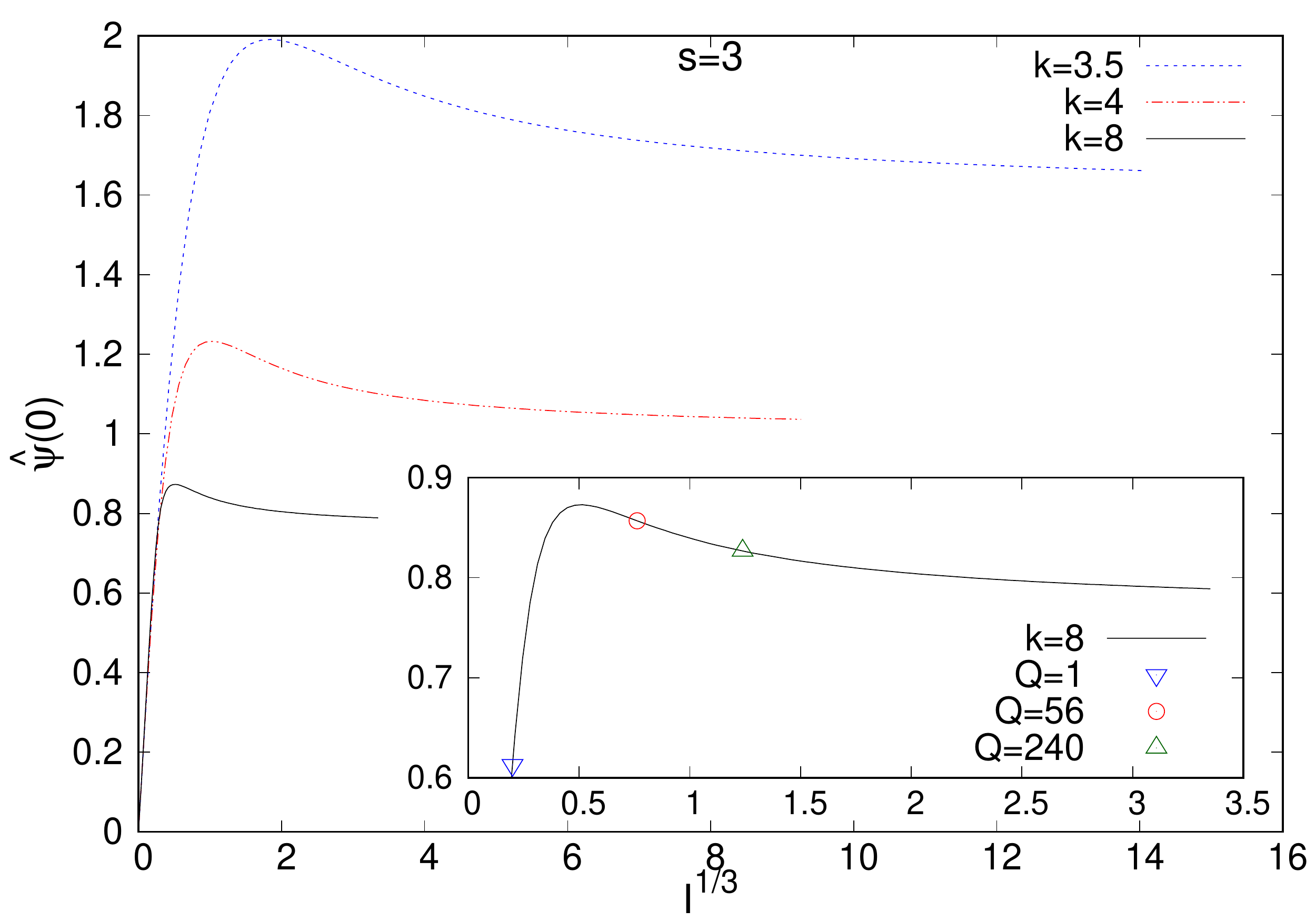}
			\caption{The value of the solution of \rf{hpsieq} at the origin, i.e. $\widehat{\psi}\(0\)$, against  $I^{1/3}$ for $s=3$ and $\kappa=3.5,\,4,\,8$. In the zoomed plot we show the values of  $\widehat{\psi}\(0\)$ for the values of baryonic charge $Q=1,\,56,\,240$,  for the particular choice of  $^{56}$Fe as the reference nuclei and the value of the ratio $\mu_0/\beta_2$ fixed by \rf{fixingbeta0mu0}. The integration constant $\gamma$ decreases from the left to the right, with its highest value being $100$. For $\kappa=3.5,\,4,\,8$ the lowest values of $\gamma$ are respectively $1.905,\,2.025,\,1.660$.}
		\label{fig:psi0}
\end{center} 
\end{figure}

In Figure \ref{fig:psi0} we show how the value of $\hpsi$ at the origin, i.e. $\hpsi\(\zeta=0\)$, varies with the quantity $I^{1/3}$, with $I$ defined in \rf{varthetaIdef}. Note that  according to \rf{chargefinal}, $I$  is proportional to the baryonic charge. From \rf{densitypsihat} we have that the baryonic charge density at the origin is $\rho\(0\) = \frac{\vartheta}{4\,\pi^2}\hpsi^s\(0\)$. The value of the integration constant $\gamma$ decreases as we move along the curves from left to right. We observe that, as $\gamma$ decreases, $\hpsi\(0\)$ increases up to a maximum and then decreases, approaching a constant value as $\gamma$ goes to its critical value (see \rf{critical}).  However, the plots in Figure \ref{fig:psi0} end before $\gamma$ reaches its critical value. For $s=3$ and $\kappa=3.5,\,4,\,8$ the lowest values of $\gamma$ are respectively $1.905,\,2.025,\,1.660$. In the zoomed plot in Figure \ref{fig:psi0}  we show, for $s=3$ and $\kappa=8$, the values of $\hpsi\(0\)$ for $Q=1$, $56$ and $240$, for the choice of $^{56}$Fe as the reference nucleus (see \rf{chargefe56}). 

The Table \ref{tables_s} below shows that the models corresponding to $s=3$ and $\kappa=6,\,...,\,20$ are  in good agreement with the experimental data for both $R$ and $E_B$,  specially for $A\geq 12$ where the RMSD of $\sqrt{\langle r^2\rangle}$ drops to about half and the RMSD of $E_B$ drops to about a quarter. The  rms radii experimental data [33] for the list of $265$ nuclei defined above, can be fitted by   
\br \sqrt{\langle r^2\rangle}_{exp.} &=& \(0.56386 + 0.839967\,Q^{1/3}\)\,fm
\nonumber\\
 \sqrt{\langle r^2\rangle}_{exp.}^{A \geq 12} &=& \(0.50530 + 0.85156\,Q^{1/3}\)\,fm
 \er
and the RMSD are respectively $0.07555\,fm$ and $0.03789\,fm$, which are close to the values of $\Delta\, R$ and  $\Delta\, R^{A \geq 12}$ for $s=3$ and $\kappa \geq 6$, shown in Table \ref{tables_s} below. Note the smooth variation of $\Delta R_{s=3}$ and $\Delta R_{s=3}^{A\geq 12}$ for $\kappa\geq 6$, being respectively about $3.0\,\%$ and $7.5\,\%$ as compared  to its maximum value at $\kappa=6$. Note also that $\Delta E_{B,{s=3}}$  monotonically decreases from its maximum value at $\kappa=7$ to its minimum value at $\kappa=20$. The situation reverses for the case  of  $\Delta E_{B,s=3}^{A \geq 12}$ which is  monotonically increasing for all $\kappa$. For $s=3$ the maximum variation of $\Delta R$, $\Delta R^{A\geq 12}$, $\Delta E_B$ and $\Delta E_B^{A\geq 12}$ in relation to its maximum for all values of $\kappa$ are respectively $36.3\,\% ,\,60.0\,\% ,\,0.5\,\% ,\,1.3\,\%$. 

\begin{center}
  \begin{longtable}{@{\extracolsep{\fill}}c c c c c c c c c@{}}   
	\caption{The RMSD $\Delta R\;\(fm\)$ of the rms radii  and $\Delta R^{A \geq 12}\;\(fm\)$ for any the integer value of $\kappa>s$ up to $\kappa=20$ and for $s=5/2,\,3,\,5$ and for $s=3$. We also give  the RMSD of the binding energy per nucleon $\Delta\,E_B\;\(MeV\)$ and $\Delta\,E_B^{A \geq 12}\;\(MeV\)$.}\label{tables_s}\\
	\hline
$\kappa$ & $\Delta R_{s=5} $ & $\Delta R^{A\geq 12}_{s=5}$ & $\Delta R_{s=5/2} $ & $\Delta R^{A\geq 12}_{s=5/2}$ & $\Delta R_{s=3} $ & $\Delta R^{A\geq 12}_{s=3}$ & $\Delta E_{B,\,{s=3}} $ & $\Delta E_{B,\,{s=3}}^{A\geq 12}$ \\ \hline
\endfirsthead
\multicolumn{9}{c}%
{\tablename\ \thetable\ -- \textit{Continued from previous page}} \\
\hline
$\kappa$ & $\Delta R_{s=5} $ & $\Delta R^{A\geq 12}_{s=5}$ & $\Delta R_{s=5/2} $ & $\Delta R^{A\geq 12}_{s=5/2}$ & $\Delta R_{s=3} $ & $\Delta R^{A\geq 12}_{s=3}$ & $\Delta E_{B,\,{s=3}} $ & $\Delta E_{B,\,{s=3}}^{A\geq 12}$ \\ \hline
\endhead
\hline \multicolumn{9}{c}{\textit{Continued on next page}} \\
\endfoot
\hline
\endlastfoot
$3$ & $*$ & $*$ & $0.63344$ & $0.61523$ & $*$ & $*$ & $*$ & $*$ \\
$4$ & $*$ & $*$ & $0.41324$ & $0.38721$ & $0.12874$ & $0.10108$ & $0.215518$ & $0.047759$ \\
$5$ & $*$ & $*$ & $0.31961$ & $0.28696$ & $0.09332$ & $0.05644$ & $0.216087$ & $0.047923$ \\
$6$ & $0.33417$ & $0.31942$ & $0.27335$ & $0.23556$ & $0.08454$ & $0.04373$ & $0.216271$ & $0.048066$ \\
$7$ & $0.29802$ & $0.28192$ & $0.24750$ & $0.20578$ & $0.08241$ & $0.04077$ & $0.216295$ & $0.048156$ \\
$8$ & $0.27233$ & $0.25519$ & $0.23165$ & $0.18693$ & $0.08204$ & $0.04047$ & $0.216247$ & $0.048213$ \\
$9$ & $0.25317$ & $0.23518$ & $0.22123$ & $0.17421$ & $0.08216$ & $0.04079$ & $0.216166$ & $0.048250$ \\
$10$ & $0.23834$ & $0.21964$ & $0.21401$ & $0.16519$ & $0.08239$ & $0.04118$ & $0.216071$ & $0.048275$ \\
$11$ & $0.22654$ & $0.20723$ & $0.20878$ & $0.15853$ & $0.08263$ & $0.04151$ & $0.215972$ & $0.048292$ \\
$12$ & $0.21693$ & $0.19709$ & $0.20487$ & $0.15347$ & $0.08284$ & $0.04177$ & $0.215873$ & $0.048306$ \\
$13$ & $0.20896$ & $0.18864$ & $0.20185$ & $0.14952$ & $0.08302$ & $0.04196$ & $0.215777$ & $0.048317$ \\
$14$ & $0.20225$ & $0.18150$ & $0.19948$ & $0.14638$ & $0.08317$ & $0.04211$ & $0.215686$ & $0.048326$ \\
$15$ & $0.19652$ & $0.17538$ & $0.19757$ & $0.14383$ & $0.08330$ & $0.04221$ & $0.215600$ & $0.048334$ \\
$16$ & $0.19157$ & $0.17009$ & $0.19601$ & $0.14173$ & $0.08341$ & $0.04229$ & $0.215520$ & $0.048342$ \\
$17$ & $0.18726$ & $0.16546$ & $0.19472$ & $0.13998$ & $0.08350$ & $0.04234$ & $0.215445$ & $0.048349$ \\
$18$ & $0.18346$ & $0.16137$ & $0.19364$ & $0.13850$ & $0.08357$ & $0.04238$ & $0.215375$ & $0.048355$ \\
$19$ & $0.18010$ & $0.15774$ & $0.19272$ & $0.13724$ & $0.08364$ & $0.04241$ & $0.215310$ & $0.048362$ \\
$20$ & $0.17711$ & $0.15450$ & $0.19193$ & $0.13616$ & $0.08369$ & $0.04243$ & $0.215249$ & $0.048368$ \\
\hline
\end{longtable}
\end{center}

\subsection{The influence of the reference nucleus on the data analysis}

We have seen that the integration constant $\gamma$ plays the role of a running coupling constant, in the sense that as we vary it, all the relevant quantities associated to the solution, including the quantity $I$ defined in \rf{varthetaIdef}, also vary. In fact, in  Figures \ref{fig:s=5_2a}, \ref{fig:s=3a} and \ref{fig:s=5a} we have shown that $I$, and consequently the baryonic charge $Q$ (see  \rf{chargefinal}), are monotonically decreasing functions of $\gamma$. In order to find what value of $\gamma$ corresponds to a given value of $Q$, we have to choose a reference nucleus and use some of its experimental  data to fix the scales. 

In \rf{fixingbeta0mu0} the value of the ratio of coupling constants $\mu_0/\beta_2$ was fixed by choosing the value of the parameter $a$ that gives the rate of the fallout of the solution at large distances. Then choosing $^{56}$Fe as the reference nucleus, and using the experimental value of its rms charge radius,  we found the value of $\gamma$ corresponding to $^{56}$Fe through \rf{radiusfe56}. Then the value of the coupling constant $\vartheta$ was fixed through \rf{chargefe56}. Once that is done the discrete sequence of values of $\gamma$ corresponding to all the other nuclei are found, using \rf{chargefinal}, through
\be
Q=\frac{Q_{\rm ref.}}{I_{\rm ref.}}\, I
\ee
where $Q_{\rm ref.}$ is the baryonic charge of the reference nucleus, and $I_{\rm ref.}=I\(\gamma_{\rm ref.}\,,\,s\,,\,\kappa\)$, with $\gamma_{\rm ref.}$ being the value of $\gamma$ found through \rf{radiusfe56}.

The question is how our data analysis is affected by the change of the reference nucleus. Clearly, if the ratio $\frac{Q_{\rm ref.}}{I_{\rm ref.}}$ is not affected much by such a change, nothing will be drastically modified. Indeed, the value of $I$ corresponding to a given $Q$ will  change little, and so it will not change drastically the value of $\gamma$ associated to that same $Q$, since $I$ is a monotonic (decreasing) function of $\gamma$.  

\begin{figure}[H]
\begin{center}
		\includegraphics[scale=0.6]{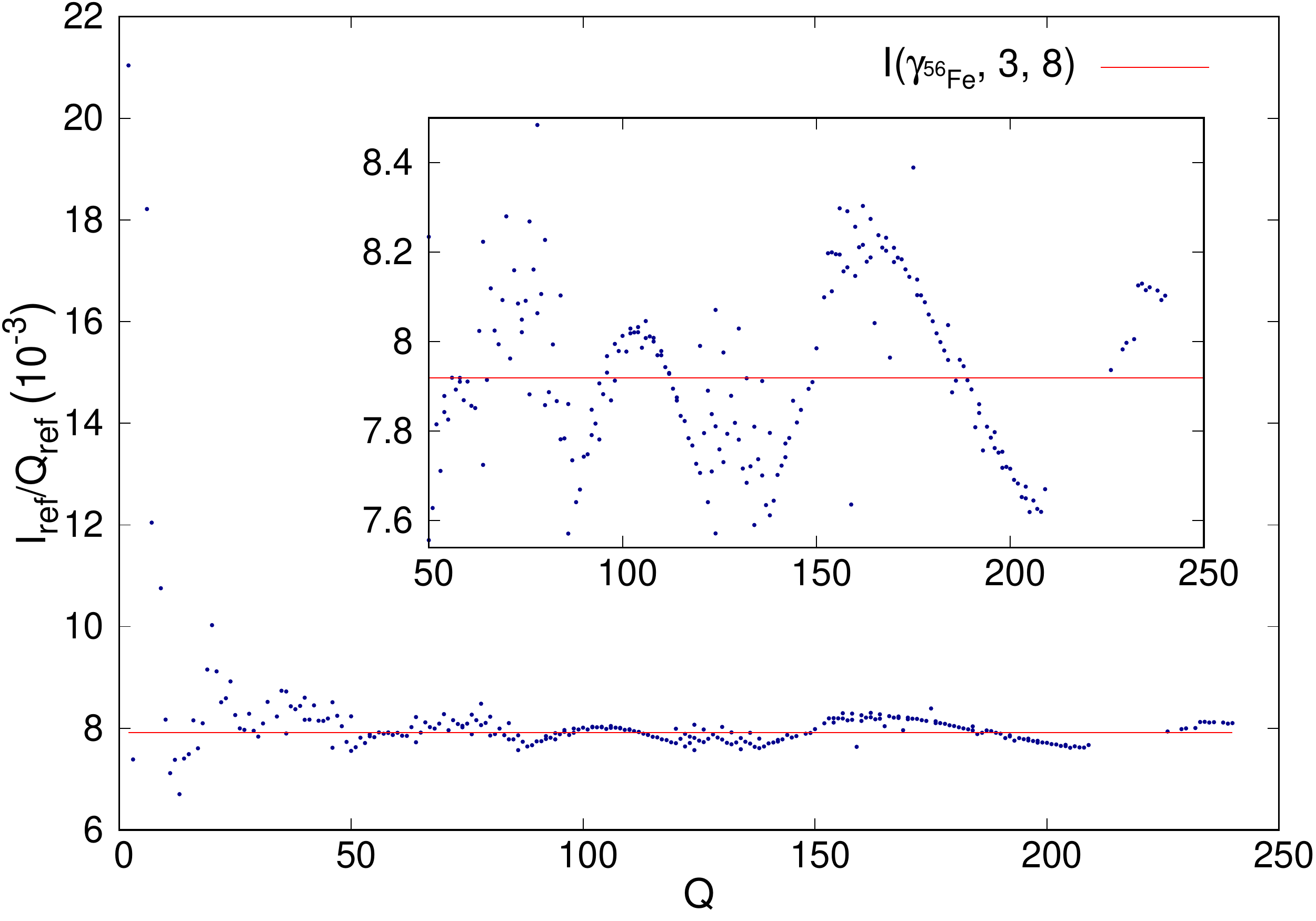}
		\caption{The $I_{{\rm ref.}}\(\gamma_{{\rm ref.}},\,s,\,\kappa\)/Q_{{\rm ref.}}$ corresponding to taking each nuclei in our list as the referential nuclei vs. $Q$ for $s=3$ and $\kappa=8$, but without distinguishing the isotopes.}
		\label{fig:desvio}
\end{center} 
\end{figure}

In order to give some examples, let us fix $\mu_0/\beta_2$ as in \rf{fixingbeta0mu0}. The choice of $^{56}$Fe as the reference nucleus leads to a RMSD of the radii of $0.00820\,fm$ and $I\(\gamma\(^{56}{\rm Fe}\),\,\kappa=8,\,s=3\)/56=0.0079194$. If we take instead $^{108}$Pd as the reference nucleus we get a very similar ratio, i.e. $I\(\gamma\(^{108}{\rm Pd}\),\,\kappa=8,\,s=3\)/108=0.0080002$, and as expected we obtain a similar RMSD of the radii given by $0.08225\,fm$. Taking the reference nucleus as any of the nuclei with $Q>50$ in our list defined at the beginning of section \ref{sec:data},  the ratio $I\(\gamma_{{\rm ref.}},\,\kappa=8,\,s=3\)/Q_{{\rm ref.}}$ oscillates between a minimum of $0.0075710$ at $^{86}$Kr and a maximum of $0.0084839$ at $^{78}$Kr. In Figure \ref{fig:desvio} we plot the ratio $I\(\gamma_{{\rm ref.}},\,\kappa=8,\,s=3\)/Q_{{\rm ref.}}$ against $Q$, by taking  each one of the $265$ nuclei in our list mentioned above, as the reference nucleus.  Therefore,  we should not expected a drastic change in the plots of Figures \ref{fig:radii} and \ref{fig:binding} for the radii and binding energies, by  replacing  $^{56}$Fe by another heavy reference nucleus in our list.  The only restriction we have is the following. As we have seen in \rf{radiusbiggamma}, and also in the Figures \ref{fig:s=2.5}, \ref{fig:s=3} and \ref{fig:s=5}, our model presents a lower bound for the radius for $\gamma\rightarrow \infty$. Therefore, a restriction on the choice of the reference nucleus is that it can not have an experimental rms radius  $R^{\rm exp}$ lower than that bound. Using \rf{fixingbeta0mu0} for $s=3$ we get from \rf{radiusbiggamma}  that $\sqrt{\langle r^2\rangle}_{{\rm min.}}= 1.8195\,fm$, and so we cannot take as the reference nucleus only $^1$H and $^3$He in our list of $N_c=265$ nuclei defined at the beginning of section \ref{sec:data}.
 
\begin{figure}[H]
\begin{center}
		\includegraphics[scale=0.6]{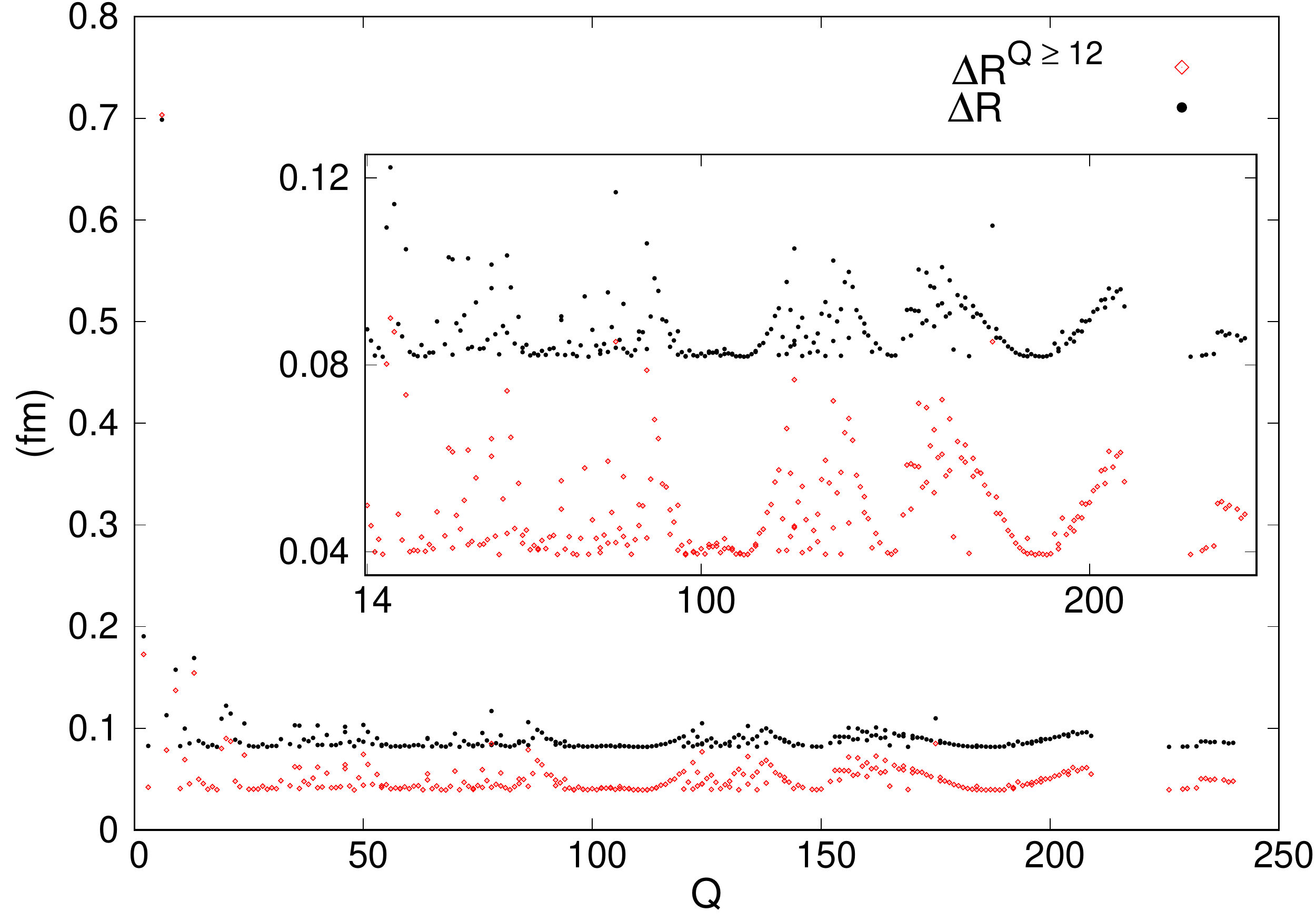}
		\caption{The smallest $\Delta R$ and $\Delta R^{A\geq 12}$ for $s=3$ among all values of $\kappa=4,\,...,\,15$ for each choice of the reference nucleus in our list, defined at the beginning of section \ref{sec:data}, excluding $^1$H and $^4$He.}
		\label{fig:rmsd}
\end{center} 
\end{figure}

  For $s=3$, and $\kappa=4,\,...,\,14$, we have evaluated the radii RMSD for every choice of the reference nucleus in our list, except for the $^1$H and $^3$He\footnote{The step size of the Runge-Kutta was reduced to $h=0.005$.}. We have obtained the values of $\kappa$ that gives the lowest values of $\Delta R$ and $\Delta R^{A \geq 12}$, as defined in \rf{rmsd}. Such RMSD values are plotted in the Figure \ref{fig:rmsd}, without distinguishing the isotopes, and we see that it does not change drastically  for non-light nuclei. So our data analysis should not change much by the replacement of the reference nucleus.     In addition, we obtain that:
\begin{enumerate}
\item The smallest $\Delta R$ occurs for $^{18}$O at $\kappa=7$ where $\Delta R=0.0818\,fm$ and $\Delta R^{A\geq 12}=0.0395\,fm$.
\item The smallest $\Delta R^{A\geq 12}$ occurs for $^{68}$Zn at $\kappa=6$ where $\Delta R=0.0825\,fm$ and $\Delta R^{A\geq 12}=0.0394\,fm$. 
\end{enumerate}   

\section{Conclusion}
\label{sec:conclusion}
\setcounter{equation}{0}

The Skyrme-type model presented in this paper reproduces with quite good accuracy, in a simple and robust way, some bulk properties of nuclei, like the radii and the binding energies, for a very wide range of values of the mass number. It is a very  good improvement as compared to the results obtained by other modifications of the Skyrme model, and may be of interest for applications in other areas of physics too. 

Our results are better for heavier nuclei than for light ones, and that is expectable since our density of baryonic charge is spherically symmetric mainly due to Coleman's false vacuum arguments. One could improve on that in several ways, including the consideration of other potentials. However, we have observed that the RMSD for radii and binding energies are insensitive to $\kappa$ as long as $s=3$ and $\kappa\geq 6$. We believe that the first improvement of the model should involve the breaking of the self-duality, which could be achieved  by introducing $h$-dependent terms like kinetic and potential energies for the $h$-fields. That modification of the model has to be small and could perhaps be treated by a perturbation theory. It could lead to non-spherically symmetric solutions and perhaps to a better description of light nuclei. 

Another line to be pursued is the study of time dependent (spinning) solutions,  that could break even more the symmetries, as for instance the degeneracy between protons and neutrons, and enrich the spectrum of the model. In order to do that one needs an  action in Minkowski space-time. The action corresponding to $E_1$ is given in \cite{laf2017,us}, and a possible action associated to $E_2$ is
\be
S_2= \int d^4x\,\left[\frac{\mu_0^2}{4\,s^2}\,K^{-2\(1-1/s\)}\,C_{\mu\nu}^2-V\(K\)+G\(U\)\,K\right] 
\ee
with  
\be
B_{\mu}=-\frac{i}{12\,\lambda^3}\,\ve_{\mu\nu\rho\sigma}\trace\(R^{\nu}\,R^{\rho}\,R^{\sigma}\); \qquad \quad C_{\mu\nu}=\partial_{\mu}B_{\nu}-\partial_{\nu}B_{\mu}; \qquad \quad K=\sqrt{B_{\mu}\,B^{\mu}}
\ee
 Of course that should be considered as a low energy effective action, and one should restrict to low frequencies where $B_0^2>B_i^2$. 

But the most interesting question to address is that of the nature of such a model. Since it reproduces some bulk properties of nuclei, it would be interesting to investigate if it could give some insights into the strong coupled regime of the strong interactions and if it connects to  some low energy limit of QCD. 

\vspace{2cm}

{\bf Acknowledgements:} LAF is partially supported by Conselho Nacional de Desenvolvimento Cient\'ifico e Tecnol\'ogico - CNPq (contract 308894/2018-9), and LRL is supported by a CAPES scholarship. 

\newpage

\appendix

\section{The numerical methods}
\label{sec:numerical}
\setcounter{equation}{0}

The solutions of the ordinary differential equation \rf{hpsieq}, satisfying $\hpsi'\(0\)=0$ and $\hpsi\(\infty\)=\hpsi'\(\infty\)=0$, were obtained, for each value of $\kappa$ and $s$, using the fourth-order explicit Runge-Kutta method with the step size $\Delta \zeta=10^{-4}$, and $\zeta$ lying in the finite interval $\left[0,\zeta_{\rm max.}\right]$. The value of $\zeta_{\rm max}$, and so the size of the adaptive lattice, is the maximum value of $\zeta$ such that $0<\hpsi\(\zeta_{\rm max.}\)<1.2\times 10^{-5}$, $\hpsi\(\zeta_{\rm max.}\)'<0$ and $\left| \hpsi'\(\zeta_{\rm max.}\) \right| <10^{-8}$. We use the shooting method by varying the value of $\hpsi\(0\)$, and our final numerical solution will be an undershoot configuration approximating the bounce solution in the finite interval given above, such that any increment of $10^{-14}$ in $\hpsi\(0\)$ gives an overshoot configuration. Typically we have $\zeta_{\rm max.} \sim 10-20$. 

The discrete sequence of values of $\gamma$ corresponding to integer values of the baryonic charge $Q$ is found in two steps. First, we obtain the value of  $\gamma\({^{56}{\rm Fe}}\)$ successively solving the equation \rf{hpsieq} by varying $\gamma$ using unbounded binary search until $\sqrt{J/I}$ differs from its value given by the relation \rf{radiusfe56}, 
by less than $10^{-10}$, where $3.7377\, fm$ is the value of the rms radius of the nucleus ${^{56}{\rm Fe}}$, according to [33]. The value of $\vartheta$ is fixed through the equation \rf{chargefe56}.  

Second, we do the same to find the discrete sequence of $\gamma$ corresponding to the integer values of the Skyrme charges $Q$, given by \rf{chargefinal}, where the ratio $\frac{\mu_0}{\beta_2}$ is fixed through \rf{fixingbeta0mu0}, which can be rewritten as
\be 
Q  = \frac{I\(\gamma\,,\,s\,,\,\kappa\)}{I\(\gamma\({^{56}{\rm Fe}}\)\,,\,s\,,\,\kappa\)/56}
\ee
such that any of its numerical value $Q_{{\rm num.}}$ differs from the nearest integer value of $Q$ by $\mid (Q-Q_{{\rm num.}})/Q\mid< 2\times 10^{-8}$.

Once this two steps are done, the bounce solution $\hpsi\(\zeta\)$ is  fixed for each value of $Q$, and so the corresponding Skyrme charge density $\rho$  is determined  by \rf{densitypsihat}.

\section{The experimental and numerical data}
\label{sec:longtable}
\setcounter{equation}{0}

In the Table \ref{tables_data} below, we provide the experimental and numerical data used in the plots of Figures \ref{fig:radii} and \ref{fig:binding}. $A$ is the mass number  and $Z$ is the number of protons of the nucleus. $R^{{\rm exp.}}$ stands for the experimental values of the rms charge radii as given in reference [33], and $E_B^{{\rm exp.}}$ stands for the experimental values of the binding energy per nucleon as given in [34].
We also provide the numerical results (with the same number of digits as the experimental data) obtained from our model for the case $s=3$ and $\kappa=8$. $R^{{\rm num.}}$ stands for the numerical rms radii, and $E_B^{{\rm num.}}$ for the numerical binding energy per nucleon. We also give the values of the parameter $\gamma$ for each value of $A$, i.e. those that lead to an integer value of the topological charge.  Since there is more than one stable nucleus for some values of $A$, but there is only one solution for each Skyrme charge $Q$, we repeat the numerical values in those cases for easy comparison.

\begin{center}
  \begin{longtable}{@{\extracolsep{\fill}}c c c c c c c c@{}} 
\caption{The experimental data $R^{{\rm exp.}}\,(fm)$  and $E_B^{{\rm exp.}}\,(MeV)$ and the numerical values $R^{{\rm num.}}\,(fm)$, $E_B^{{\rm num.}}\,(MeV)$ and $\gamma$ (dimensionless) obtained from our model  for the case $s=3$ and $\kappa=8$.}\label{tables_data}\\
\hline
Name & $A$ & $Z$ & $R^{{\rm exp.}}$ & $R^{{\rm num.}}$ & $E_B^{{\rm exp.}}$ & $E_B^{{\rm num.}}$ & $\gamma$\\
\hline
\endfirsthead
\multicolumn{8}{c}%
{\tablename\ \thetable\ -- \textit{Continued from previous page}} \\
\hline
Name & $A$ & $Z$ & $R^{{\rm exp.}}$ & $R^{{\rm num.}}$ & $E_B^{{\rm exp.}}$ & $E_B^{{\rm num.}}$ & $\gamma$\\
\hline
\endhead
\hline \multicolumn{8}{c}{\textit{Continued on next page}} \\
\endfoot
\hline
\endlastfoot
H & $ 1 $ & $ 1 $ & $ 0.8783 $ & $ 1.8467 $ & $ 0.000000 $ & $ 0.000000 $ & $ 4.491068461142762 $ \\
H & $ 2 $ & $ 1 $ & $ 2.1421 $ & $ 1.9087 $ & $ 1.112283 $ & $ 2.895106 $ & $ 3.649869712047377 $ \\
He & $ 3 $ & $ 2 $ & $ 1.9661 $ & $ 1.9808 $ & $ 2.572680 $ & $ 4.254723 $ & $ 3.275512965240602 $ \\
He & $ 4 $ & $ 2 $ & $ 1.6755 $ & $ 2.0527 $ & $ 7.073915 $ & $ 5.084109 $ & $ 3.055186690308611 $ \\
Li & $ 6 $ & $ 3 $ & $ 2.5890 $ & $ 2.1862 $ & $ 5.332331 $ & $ 6.078895 $ & $ 2.797917904594600 $ \\
Li & $ 7 $ & $ 3 $ & $ 2.4440 $ & $ 2.2473 $ & $ 5.606439 $ & $ 6.407667 $ & $ 2.714002952968850 $ \\
Be & $ 9 $ & $ 4 $ & $ 2.5190 $ & $ 2.3598 $ & $ 6.462668 $ & $ 6.889732 $ & $ 2.591114911251203 $ \\
B & $ 10 $ & $ 5 $ & $ 2.4277 $ & $ 2.4117 $ & $ 6.475083 $ & $ 7.072907 $ & $ 2.544219486469452 $ \\
B & $ 11 $ & $ 5 $ & $ 2.4060 $ & $ 2.4613 $ & $ 6.927732 $ & $ 7.229382 $ & $ 2.503942224117655 $ \\
C & $ 12 $ & $ 6 $ & $ 2.4702 $ & $ 2.5086 $ & $ 7.680144 $ & $ 7.364775 $ & $ 2.468850343952055 $ \\
C & $ 13 $ & $ 6 $ & $ 2.4614 $ & $ 2.5540 $ & $ 7.469849 $ & $ 7.483176 $ & $ 2.437910925143941 $ \\
N & $ 14 $ & $ 7 $ & $ 2.5582 $ & $ 2.5976 $ & $ 7.475614 $ & $ 7.587648 $ & $ 2.410358116708282 $ \\
N & $ 15 $ & $ 7 $ & $ 2.6058 $ & $ 2.6397 $ & $ 7.699460 $ & $ 7.680532 $ & $ 2.385610494137132 $ \\
O & $ 16 $ & $ 8 $ & $ 2.6991 $ & $ 2.6802 $ & $ 7.976206 $ & $ 7.763651 $ & $ 2.363217640071331 $ \\
O & $ 17 $ & $ 8 $ & $ 2.6932 $ & $ 2.7194 $ & $ 7.750728 $ & $ 7.838453 $ & $ 2.342824482983545 $ \\
O & $ 18 $ & $ 8 $ & $ 2.7726 $ & $ 2.7574 $ & $ 7.767097 $ & $ 7.906102 $ & $ 2.324146819327975 $ \\
F & $ 19 $ & $ 9 $ & $ 2.8976 $ & $ 2.7942 $ & $ 7.779018 $ & $ 7.967544 $ & $ 2.306954100038247 $ \\
Ne & $ 20 $ & $ 10 $ & $ 3.0055 $ & $ 2.8300 $ & $ 8.032240 $ & $ 8.023561 $ & $ 2.291057065022768 $ \\
Ne & $ 21 $ & $ 10 $ & $ 2.9695 $ & $ 2.8648 $ & $ 7.971713 $ & $ 8.074802 $ & $ 2.276298691132936 $ \\
Ne & $ 22 $ & $ 10 $ & $ 2.9525 $ & $ 2.8986 $ & $ 8.080465 $ & $ 8.121814 $ & $ 2.262547453280112 $ \\
Na & $ 23 $ & $ 11 $ & $ 2.9936 $ & $ 2.9316 $ & $ 8.111493 $ & $ 8.165058 $ & $ 2.249692231249988 $ \\
Mg & $ 24 $ & $ 12 $ & $ 3.0570 $ & $ 2.9638 $ & $ 8.260709 $ & $ 8.204929 $ & $ 2.237638407479905 $ \\
Mg & $ 25 $ & $ 12 $ & $ 3.0284 $ & $ 2.9952 $ & $ 8.223502 $ & $ 8.241766 $ & $ 2.226304840103443 $ \\
Mg & $ 26 $ & $ 12 $ & $ 3.0337 $ & $ 3.0259 $ & $ 8.333870 $ & $ 8.275862 $ & $ 2.215621488325438 $ \\
Al & $ 27 $ & $ 13 $ & $ 3.0610 $ & $ 3.0560 $ & $ 8.331553 $ & $ 8.307472 $ & $ 2.205527530232682 $ \\
Si & $ 28 $ & $ 14 $ & $ 3.1224 $ & $ 3.0854 $ & $ 8.447744 $ & $ 8.336819 $ & $ 2.195969856721151 $ \\
Si & $ 29 $ & $ 14 $ & $ 3.1176 $ & $ 3.1141 $ & $ 8.448635 $ & $ 8.364099 $ & $ 2.186901855810426 $ \\
Si & $ 30 $ & $ 14 $ & $ 3.1336 $ & $ 3.1424 $ & $ 8.520654 $ & $ 8.389484 $ & $ 2.178282423400496 $ \\
P & $ 31 $ & $ 15 $ & $ 3.1889 $ & $ 3.1700 $ & $ 8.481167 $ & $ 8.413128 $ & $ 2.170075152240669 $ \\
S & $ 32 $ & $ 16 $ & $ 3.2611 $ & $ 3.1972 $ & $ 8.493129 $ & $ 8.435168 $ & $ 2.162247662359158 $ \\
S & $ 34 $ & $ 16 $ & $ 3.2847 $ & $ 3.2500 $ & $ 8.583498 $ & $ 8.474908 $ & $ 2.147619395869828 $ \\
Cl & $ 35 $ & $ 17 $ & $ 3.3654 $ & $ 3.2758 $ & $ 8.520278 $ & $ 8.492817 $ & $ 2.140769427267332 $ \\
S & $ 36 $ & $ 16 $ & $ 3.2985 $ & $ 3.3011 $ & $ 8.575389 $ & $ 8.509539 $ & $ 2.134200134292854 $ \\
Ar & $ 36 $ & $ 18 $ & $ 3.3905 $ & $ 3.3011 $ & $ 8.519909 $ & $ 8.509539 $ & $ 2.134200134292854 $ \\
Cl & $ 37 $ & $ 17 $ & $ 3.3840 $ & $ 3.3260 $ & $ 8.570281 $ & $ 8.525155 $ & $ 2.127892515622490 $ \\
Ar & $ 38 $ & $ 18 $ & $ 3.4028 $ & $ 3.3505 $ & $ 8.614280 $ & $ 8.539738 $ & $ 2.121829333480677 $ \\
K & $ 39 $ & $ 19 $ & $ 3.4349 $ & $ 3.3746 $ & $ 8.557025 $ & $ 8.553355 $ & $ 2.115994908123602 $ \\
Ca & $ 40 $ & $ 20 $ & $ 3.4776 $ & $ 3.3983 $ & $ 8.551303 $ & $ 8.566065 $ & $ 2.110374940891559 $ \\
Ar & $ 40 $ & $ 18 $ & $ 3.4274 $ & $ 3.3983 $ & $ 8.595259 $ & $ 8.566065 $ & $ 2.110374940891559 $ \\
K & $ 41 $ & $ 19 $ & $ 3.4518 $ & $ 3.4217 $ & $ 8.576072 $ & $ 8.577924 $ & $ 2.104956361248827 $ \\
Ca & $ 42 $ & $ 20 $ & $ 3.5081 $ & $ 3.4448 $ & $ 8.616563 $ & $ 8.588984 $ & $ 2.099727194058634 $ \\
Ca & $ 43 $ & $ 20 $ & $ 3.4954 $ & $ 3.4675 $ & $ 8.600663 $ & $ 8.599289 $ & $ 2.094676444000906 $ \\
Ca & $ 44 $ & $ 20 $ & $ 3.5179 $ & $ 3.4899 $ & $ 8.658175 $ & $ 8.608884 $ & $ 2.089793994574852 $ \\
Sc & $ 45 $ & $ 21 $ & $ 3.5459 $ & $ 3.5120 $ & $ 8.618931 $ & $ 8.617808 $ & $ 2.085070519558121 $ \\
Ca & $ 46 $ & $ 20 $ & $ 3.4953 $ & $ 3.5338 $ & $ 8.66898 $ & $ 8.62610 $ & $ 2.080497405145488 $ \\
Ti & $ 46 $ & $ 22 $ & $ 3.6070 $ & $ 3.5338 $ & $ 8.656451 $ & $ 8.626097 $ & $ 2.080497405145488 $ \\
Ti & $ 47 $ & $ 22 $ & $ 3.5962 $ & $ 3.5554 $ & $ 8.661227 $ & $ 8.633785 $ & $ 2.076066681276107 $ \\
Ti & $ 48 $ & $ 22 $ & $ 3.5921 $ & $ 3.5766 $ & $ 8.723006 $ & $ 8.640903 $ & $ 2.071770960893693 $ \\
Ti & $ 49 $ & $ 22 $ & $ 3.5733 $ & $ 3.5976 $ & $ 8.711157 $ & $ 8.647481 $ & $ 2.067603386077977 $ \\
Ti & $ 50 $ & $ 22 $ & $ 3.5704 $ & $ 3.6183 $ & $ 8.755718 $ & $ 8.653545 $ & $ 2.063557580146583 $ \\
Cr & $ 50 $ & $ 24 $ & $ 3.6588 $ & $ 3.6183 $ & $ 8.701032 $ & $ 8.653545 $ & $ 2.063557580146583 $ \\
V & $ 51 $ & $ 23 $ & $ 3.6002 $ & $ 3.6388 $ & $ 8.742099 $ & $ 8.659121 $ & $ 2.059627604959952 $ \\
Cr & $ 52 $ & $ 24 $ & $ 3.6452 $ & $ 3.6590 $ & $ 8.775989 $ & $ 8.664232 $ & $ 2.055807922773732 $ \\
Cr & $ 53 $ & $ 24 $ & $ 3.6511 $ & $ 3.6790 $ & $ 8.760198 $ & $ 8.668900 $ & $ 2.052093362076591 $ \\
Cr & $ 54 $ & $ 24 $ & $ 3.6885 $ & $ 3.6988 $ & $ 8.777955 $ & $ 8.673146 $ & $ 2.048479086930095 $ \\
Fe & $ 54 $ & $ 26 $ & $ 3.6933 $ & $ 3.6988 $ & $ 8.736382 $ & $ 8.673146 $ & $ 2.048479086930095 $ \\
Mn & $ 55 $ & $ 25 $ & $ 3.7057 $ & $ 3.7184 $ & $ 8.765022 $ & $ 8.676988 $ & $ 2.044960569393613 $ \\
Fe & $ 56 $ & $ 26 $ & $ 3.7377 $ & $ 3.7377 $ & $ 8.790354 $ & $ 8.680445 $ & $ 2.041533564673800 $ \\
Fe & $ 57 $ & $ 26 $ & $ 3.7532 $ & $ 3.7568 $ & $ 8.770279 $ & $ 8.683533 $ & $ 2.038194088685605 $ \\
Fe & $ 58 $ & $ 26 $ & $ 3.7745 $ & $ 3.7758 $ & $ 8.792250 $ & $ 8.686269 $ & $ 2.034938397753139 $ \\
Ni & $ 58 $ & $ 28 $ & $ 3.7757 $ & $ 3.7758 $ & $ 8.732059 $ & $ 8.686269 $ & $ 2.034938397753139 $ \\
Co & $ 59 $ & $ 27 $ & $ 3.7875 $ & $ 3.7945 $ & $ 8.768035 $ & $ 8.688667 $ & $ 2.031762970212437 $ \\
Ni & $ 60 $ & $ 28 $ & $ 3.8118 $ & $ 3.8130 $ & $ 8.780774 $ & $ 8.690741 $ & $ 2.028664489709483 $ \\
Ni & $ 61 $ & $ 28 $ & $ 3.8225 $ & $ 3.8314 $ & $ 8.765025 $ & $ 8.692506 $ & $ 2.025639830011463 $ \\
Ni & $ 62 $ & $ 28 $ & $ 3.8399 $ & $ 3.8495 $ & $ 8.794553 $ & $ 8.693972 $ & $ 2.022686041171518 $ \\
Cu & $ 63 $ & $ 29 $ & $ 3.8823 $ & $ 3.8675 $ & $ 8.752138 $ & $ 8.695152 $ & $ 2.019800336906873 $ \\
Ni & $ 64 $ & $ 28 $ & $ 3.8572 $ & $ 3.8853 $ & $ 8.777461 $ & $ 8.696057 $ & $ 2.016980083066171 $ \\
Zn & $ 64 $ & $ 30 $ & $ 3.9283 $ & $ 3.8853 $ & $ 8.735905 $ & $ 8.696057 $ & $ 2.016980083066171 $ \\
Cu & $ 65 $ & $ 29 $ & $ 3.9022 $ & $ 3.9029 $ & $ 8.757096 $ & $ 8.696698 $ & $ 2.014222787076635 $ \\
Zn & $ 66 $ & $ 30 $ & $ 3.9491 $ & $ 3.9204 $ & $ 8.759632 $ & $ 8.697084 $ & $ 2.011526088274126 $ \\
Zn & $ 67 $ & $ 30 $ & $ 3.9530 $ & $ 3.9377 $ & $ 8.734152 $ & $ 8.697226 $ & $ 2.008887749029717 $ \\
Zn & $ 68 $ & $ 30 $ & $ 3.9658 $ & $ 3.9549 $ & $ 8.755680 $ & $ 8.697131 $ & $ 2.006305646596767 $ \\
Ga & $ 69 $ & $ 31 $ & $ 3.9973 $ & $ 3.9719 $ & $ 8.724579 $ & $ 8.696810 $ & $ 2.003777765609718 $ \\
Ge & $ 70 $ & $ 32 $ & $ 4.0414 $ & $ 3.9887 $ & $ 8.721700 $ & $ 8.696268 $ & $ 2.001302191174485 $ \\
Ga & $ 71 $ & $ 31 $ & $ 4.0118 $ & $ 4.0054 $ & $ 8.717604 $ & $ 8.695516 $ & $ 1.998877102495371 $ \\
Ge & $ 72 $ & $ 32 $ & $ 4.0576 $ & $ 4.0220 $ & $ 8.731745 $ & $ 8.694559 $ & $ 1.996500766990397 $ \\
Ge & $ 73 $ & $ 32 $ & $ 4.0632 $ & $ 4.0384 $ & $ 8.705049 $ & $ 8.693404 $ & $ 1.994171534851110 $ \\
Ge & $ 74 $ & $ 32 $ & $ 4.0742 $ & $ 4.0547 $ & $ 8.725200 $ & $ 8.692059 $ & $ 1.991887834007733 $ \\
Se & $ 74 $ & $ 34 $ & $ 4.0700 $ & $ 4.0547 $ & $ 8.687715 $ & $ 8.692059 $ & $ 1.991887834007733 $ \\
As & $ 75 $ & $ 33 $ & $ 4.0968 $ & $ 4.0708 $ & $ 8.700874 $ & $ 8.690530 $ & $ 1.989648165464452 $ \\
Se & $ 76 $ & $ 34 $ & $ 4.1395 $ & $ 4.0868 $ & $ 8.711477 $ & $ 8.688822 $ & $ 1.987451098972989 $ \\
Ge & $ 76 $ & $ 32 $ & $ 4.0811 $ & $ 4.0868 $ & $ 8.705236 $ & $ 8.688822 $ & $ 1.987451098972989 $ \\
Se & $ 77 $ & $ 34 $ & $ 4.1395 $ & $ 4.1027 $ & $ 8.694690 $ & $ 8.686942 $ & $ 1.985295269015788 $ \\
Se & $ 78 $ & $ 34 $ & $ 4.1406 $ & $ 4.1184 $ & $ 8.717806 $ & $ 8.684895 $ & $ 1.983179371073147 $ \\
Kr & $ 78 $ & $ 36 $ & $ 4.2038 $ & $ 4.1184 $ & $ 8.661238 $ & $ 8.684895 $ & $ 1.983179371073147 $ \\
Br & $ 79 $ & $ 35 $ & $ 4.1629 $ & $ 4.1341 $ & $ 8.687594 $ & $ 8.682686 $ & $ 1.981102158150497 $ \\
Kr & $ 80 $ & $ 36 $ & $ 4.1970 $ & $ 4.1496 $ & $ 8.692928 $ & $ 8.680320 $ & $ 1.979062437544858 $ \\
Se & $ 80 $ & $ 34 $ & $ 4.1400 $ & $ 4.1496 $ & $ 8.710813 $ & $ 8.680320 $ & $ 1.979062437544858 $ \\
Br & $ 81 $ & $ 35 $ & $ 4.1599 $ & $ 4.1650 $ & $ 8.695946 $ & $ 8.677802 $ & $ 1.977059067831450 $ \\
Kr & $ 82 $ & $ 36 $ & $ 4.1919 $ & $ 4.1802 $ & $ 8.710675 $ & $ 8.675137 $ & $ 1.975090956052185 $ \\
Kr & $ 83 $ & $ 36 $ & $ 4.1871 $ & $ 4.1954 $ & $ 8.695729 $ & $ 8.672330 $ & $ 1.973157055091182 $ \\
Kr & $ 84 $ & $ 36 $ & $ 4.1884 $ & $ 4.2104 $ & $ 8.717446 $ & $ 8.669383 $ & $ 1.971256361222048 $ \\
Sr & $ 84 $ & $ 38 $ & $ 4.2394 $ & $ 4.2104 $ & $ 8.677512 $ & $ 8.669383 $ & $ 1.971256361222048 $ \\
Rb & $ 85 $ & $ 37 $ & $ 4.2036 $ & $ 4.2253 $ & $ 8.697441 $ & $ 8.666302 $ & $ 1.969387911814059 $ \\
Kr & $ 86 $ & $ 36 $ & $ 4.1835 $ & $ 4.2402 $ & $ 8.712029 $ & $ 8.663091 $ & $ 1.967550783185259 $ \\
Sr & $ 86 $ & $ 38 $ & $ 4.2307 $ & $ 4.2402 $ & $ 8.708456 $ & $ 8.663091 $ & $ 1.967550783185259 $ \\
Sr & $ 87 $ & $ 38 $ & $ 4.2249 $ & $ 4.2549 $ & $ 8.705236 $ & $ 8.659752 $ & $ 1.965744088591193 $ \\
Sr & $ 88 $ & $ 38 $ & $ 4.2240 $ & $ 4.2695 $ & $ 8.732595 $ & $ 8.656290 $ & $ 1.963966976339631 $ \\
Y & $ 89 $ & $ 39 $ & $ 4.2430 $ & $ 4.2840 $ & $ 8.713978 $ & $ 8.652708 $ & $ 1.962218628021685 $ \\
Zr & $ 90 $ & $ 40 $ & $ 4.2694 $ & $ 4.2984 $ & $ 8.709969 $ & $ 8.649010 $ & $ 1.960498256851153 $ \\
Zr & $ 91 $ & $ 40 $ & $ 4.2845 $ & $ 4.3127 $ & $ 8.693314 $ & $ 8.645198 $ & $ 1.958805106104303 $ \\
Zr & $ 92 $ & $ 40 $ & $ 4.3057 $ & $ 4.3269 $ & $ 8.692678 $ & $ 8.641275 $ & $ 1.957138447652745 $ \\
Mo & $ 92 $ & $ 42 $ & $ 4.3151 $ & $ 4.3269 $ & $ 8.657730 $ & $ 8.641275 $ & $ 1.957138447652745 $ \\
Nb & $ 93 $ & $ 41 $ & $ 4.3240 $ & $ 4.3410 $ & $ 8.664184 $ & $ 8.637246 $ & $ 1.955497580583371 $ \\
Zr & $ 94 $ & $ 40 $ & $ 4.3320 $ & $ 4.3550 $ & $ 8.666801 $ & $ 8.633112 $ & $ 1.953881829898621 $ \\
Mo & $ 94 $ & $ 42 $ & $ 4.3529 $ & $ 4.3550 $ & $ 8.662333 $ & $ 8.633112 $ & $ 1.953881829898621 $ \\
Mo & $ 95 $ & $ 42 $ & $ 4.3628 $ & $ 4.3690 $ & $ 8.648720 $ & $ 8.628876 $ & $ 1.952290545292565 $ \\
Mo & $ 96 $ & $ 42 $ & $ 4.3847 $ & $ 4.3828 $ & $ 8.653987 $ & $ 8.624541 $ & $ 1.950723099996405 $ \\
Ru & $ 96 $ & $ 44 $ & $ 4.3908 $ & $ 4.3828 $ & $ 8.609412 $ & $ 8.624541 $ & $ 1.950723099996405 $ \\
Mo & $ 97 $ & $ 42 $ & $ 4.3880 $ & $ 4.3966 $ & $ 8.635092 $ & $ 8.620109 $ & $ 1.949178889690030 $ \\
Mo & $ 98 $ & $ 42 $ & $ 4.4091 $ & $ 4.4102 $ & $ 8.635168 $ & $ 8.615583 $ & $ 1.947657331474187 $ \\
Ru & $ 98 $ & $ 44 $ & $ 4.4229 $ & $ 4.4102 $ & $ 8.62031 $ & $ 8.61558 $ & $ 1.947657331474187 $ \\
Ru & $ 99 $ & $ 44 $ & $ 4.4338 $ & $ 4.4238 $ & $ 8.608712 $ & $ 8.610965 $ & $ 1.946157862899922 $ \\
Ru & $ 100 $ & $ 44 $ & $ 4.4531 $ & $ 4.4373 $ & $ 8.619359 $ & $ 8.606258 $ & $ 1.944679941051426 $ \\
Ru & $ 101 $ & $ 44 $ & $ 4.4606 $ & $ 4.4507 $ & $ 8.601365 $ & $ 8.601463 $ & $ 1.943223041678503 $ \\
Ru & $ 102 $ & $ 44 $ & $ 4.4809 $ & $ 4.4640 $ & $ 8.607427 $ & $ 8.596583 $ & $ 1.941786658376123 $ \\
Pd & $ 102 $ & $ 46 $ & $ 4.4827 $ & $ 4.4640 $ & $ 8.580290 $ & $ 8.596583 $ & $ 1.941786658376123 $ \\
Rh & $ 103 $ & $ 45 $ & $ 4.4945 $ & $ 4.4772 $ & $ 8.584192 $ & $ 8.591620 $ & $ 1.940370301807507 $ \\
Pd & $ 104 $ & $ 46 $ & $ 4.5078 $ & $ 4.4904 $ & $ 8.584848 $ & $ 8.586576 $ & $ 1.938973498968275 $ \\
Ru & $ 104 $ & $ 44 $ & $ 4.5098 $ & $ 4.4904 $ & $ 8.587399 $ & $ 8.586576 $ & $ 1.938973498968275 $ \\
Pd & $ 105 $ & $ 46 $ & $ 4.5150 $ & $ 4.5035 $ & $ 8.570650 $ & $ 8.581453 $ & $ 1.937595792489287 $ \\
Pd & $ 106 $ & $ 46 $ & $ 4.5318 $ & $ 4.5165 $ & $ 8.579992 $ & $ 8.576252 $ & $ 1.936236739975294 $ \\
Cd & $ 106 $ & $ 48 $ & $ 4.5383 $ & $ 4.5165 $ & $ 8.539048 $ & $ 8.576252 $ & $ 1.936236739975294 $ \\
Ag & $ 107 $ & $ 47 $ & $ 4.5454 $ & $ 4.5294 $ & $ 8.553900 $ & $ 8.570976 $ & $ 1.934895913377873 $ \\
Cd & $ 108 $ & $ 48 $ & $ 4.5577 $ & $ 4.5422 $ & $ 8.550019 $ & $ 8.565626 $ & $ 1.933572898399922 $ \\
Pd & $ 108 $ & $ 46 $ & $ 4.5563 $ & $ 4.5422 $ & $ 8.567023 $ & $ 8.565626 $ & $ 1.933572898399922 $ \\
Ag & $ 109 $ & $ 47 $ & $ 4.5638 $ & $ 4.5550 $ & $ 8.547915 $ & $ 8.560203 $ & $ 1.932267293930503 $ \\
Pd & $ 110 $ & $ 46 $ & $ 4.5782 $ & $ 4.5677 $ & $ 8.547162 $ & $ 8.554710 $ & $ 1.930978711507522 $ \\
Cd & $ 110 $ & $ 48 $ & $ 4.5765 $ & $ 4.5677 $ & $ 8.551275 $ & $ 8.554710 $ & $ 1.930978711507522 $ \\
Cd & $ 111 $ & $ 48 $ & $ 4.5845 $ & $ 4.5804 $ & $ 8.537079 $ & $ 8.549149 $ & $ 1.929706774807386 $ \\
Cd & $ 112 $ & $ 48 $ & $ 4.5944 $ & $ 4.5929 $ & $ 8.544730 $ & $ 8.543519 $ & $ 1.928451119159275 $ \\
Sn & $ 112 $ & $ 50 $ & $ 4.5948 $ & $ 4.5929 $ & $ 8.513618 $ & $ 8.543519 $ & $ 1.928451119159275 $ \\
In & $ 113 $ & $ 49 $ & $ 4.6010 $ & $ 4.6054 $ & $ 8.522929 $ & $ 8.537824 $ & $ 1.927211391083395 $ \\
Sn & $ 114 $ & $ 50 $ & $ 4.6099 $ & $ 4.6178 $ & $ 8.522566 $ & $ 8.532064 $ & $ 1.925987247851143 $ \\
Cd & $ 114 $ & $ 48 $ & $ 4.6087 $ & $ 4.6178 $ & $ 8.531513 $ & $ 8.532064 $ & $ 1.925987247851143 $ \\
Sn & $ 115 $ & $ 50 $ & $ 4.6148 $ & $ 4.6302 $ & $ 8.514069 $ & $ 8.526241 $ & $ 1.924778357066431 $ \\
Sn & $ 116 $ & $ 50 $ & $ 4.6250 $ & $ 4.6425 $ & $ 8.523116 $ & $ 8.520357 $ & $ 1.923584396266858 $ \\
Sn & $ 117 $ & $ 50 $ & $ 4.6302 $ & $ 4.6547 $ & $ 8.509611 $ & $ 8.514412 $ & $ 1.922405052543427 $ \\
Sn & $ 118 $ & $ 50 $ & $ 4.6393 $ & $ 4.6668 $ & $ 8.516533 $ & $ 8.508407 $ & $ 1.921240022178119 $ \\
Sn & $ 119 $ & $ 50 $ & $ 4.6438 $ & $ 4.6789 $ & $ 8.499449 $ & $ 8.502345 $ & $ 1.920089010298075 $ \\
Sn & $ 120 $ & $ 50 $ & $ 4.6519 $ & $ 4.6909 $ & $ 8.504492 $ & $ 8.496226 $ & $ 1.918951730545584 $ \\
Te & $ 120 $ & $ 52 $ & $ 4.7038 $ & $ 4.6909 $ & $ 8.477034 $ & $ 8.496226 $ & $ 1.918951730545584 $ \\
Sb & $ 121 $ & $ 51 $ & $ 4.6802 $ & $ 4.7029 $ & $ 8.482066 $ & $ 8.490052 $ & $ 1.917827904763047 $ \\
Te & $ 122 $ & $ 52 $ & $ 4.7095 $ & $ 4.7148 $ & $ 8.478140 $ & $ 8.483823 $ & $ 1.916717262692254 $ \\
Sn & $ 122 $ & $ 50 $ & $ 4.6634 $ & $ 4.7148 $ & $ 8.487907 $ & $ 8.483823 $ & $ 1.916717262692254 $ \\
Sb & $ 123 $ & $ 51 $ & $ 4.6879 $ & $ 4.7266 $ & $ 8.472328 $ & $ 8.477540 $ & $ 1.915619541686807 $ \\
Te & $ 123 $ & $ 52 $ & $ 4.7117 $ & $ 4.7266 $ & $ 8.465546 $ & $ 8.477540 $ & $ 1.915619541686807 $ \\
Te & $ 124 $ & $ 52 $ & $ 4.7183 $ & $ 4.7384 $ & $ 8.473279 $ & $ 8.471206 $ & $ 1.914534486437504 $ \\
Xe & $ 124 $ & $ 54 $ & $ 4.7661 $ & $ 4.7384 $ & $ 8.437565 $ & $ 8.471206 $ & $ 1.914534486437504 $ \\
Sn & $ 124 $ & $ 50 $ & $ 4.6735 $ & $ 4.7384 $ & $ 8.467421 $ & $ 8.471206 $ & $ 1.914534486437504 $ \\
Te & $ 125 $ & $ 52 $ & $ 4.7204 $ & $ 4.7501 $ & $ 8.458045 $ & $ 8.464820 $ & $ 1.913461848709503 $ \\
Te & $ 126 $ & $ 52 $ & $ 4.7266 $ & $ 4.7618 $ & $ 8.463248 $ & $ 8.458383 $ & $ 1.912401387091375 $ \\
Xe & $ 126 $ & $ 54 $ & $ 4.7722 $ & $ 4.7618 $ & $ 8.443541 $ & $ 8.458383 $ & $ 1.912401387091375 $ \\
I & $ 127 $ & $ 53 $ & $ 4.7500 $ & $ 4.7734 $ & $ 8.445487 $ & $ 8.451898 $ & $ 1.911352866754402 $ \\
Xe & $ 128 $ & $ 54 $ & $ 4.7774 $ & $ 4.7850 $ & $ 8.443298 $ & $ 8.445364 $ & $ 1.910316059222751 $ \\
Xe & $ 129 $ & $ 54 $ & $ 4.7775 $ & $ 4.7964 $ & $ 8.431390 $ & $ 8.438783 $ & $ 1.909290742152922 $ \\
Ba & $ 130 $ & $ 56 $ & $ 4.8283 $ & $ 4.8079 $ & $ 8.405549 $ & $ 8.432156 $ & $ 1.908276699122978 $ \\
Xe & $ 130 $ & $ 54 $ & $ 4.7818 $ & $ 4.8079 $ & $ 8.437731 $ & $ 8.432156 $ & $ 1.908276699122978 $ \\
Xe & $ 131 $ & $ 54 $ & $ 4.7808 $ & $ 4.8193 $ & $ 8.423736 $ & $ 8.425482 $ & $ 1.907273719430179 $ \\
Xe & $ 132 $ & $ 54 $ & $ 4.7859 $ & $ 4.8306 $ & $ 8.427622 $ & $ 8.418764 $ & $ 1.906281597897336 $ \\
Ba & $ 132 $ & $ 56 $ & $ 4.8303 $ & $ 4.8306 $ & $ 8.409375 $ & $ 8.418764 $ & $ 1.906281597897336 $ \\
Cs & $ 133 $ & $ 55 $ & $ 4.8041 $ & $ 4.8418 $ & $ 8.409978 $ & $ 8.412003 $ & $ 1.905300134686859 $ \\
Xe & $ 134 $ & $ 54 $ & $ 4.7899 $ & $ 4.8531 $ & $ 8.413699 $ & $ 8.405198 $ & $ 1.904329135122647 $ \\
Ba & $ 134 $ & $ 56 $ & $ 4.8322 $ & $ 4.8531 $ & $ 8.408171 $ & $ 8.405198 $ & $ 1.904329135122647 $ \\
Ba & $ 135 $ & $ 56 $ & $ 4.8294 $ & $ 4.8642 $ & $ 8.397533 $ & $ 8.398350 $ & $ 1.903368409519181 $ \\
Ce & $ 136 $ & $ 58 $ & $ 4.8739 $ & $ 4.8753 $ & $ 8.373760 $ & $ 8.391462 $ & $ 1.902417773017244 $ \\
Ba & $ 136 $ & $ 56 $ & $ 4.8334 $ & $ 4.8753 $ & $ 8.402755 $ & $ 8.391462 $ & $ 1.902417773017244 $ \\
Ba & $ 137 $ & $ 56 $ & $ 4.8314 $ & $ 4.8864 $ & $ 8.391827 $ & $ 8.384532 $ & $ 1.901477045426730 $ \\
Ba & $ 138 $ & $ 56 $ & $ 4.8378 $ & $ 4.8974 $ & $ 8.393420 $ & $ 8.377562 $ & $ 1.900546051075015 $ \\
Ce & $ 138 $ & $ 58 $ & $ 4.8737 $ & $ 4.8974 $ & $ 8.37708 $ & $ 8.37756 $ & $ 1.900546051075015 $ \\
La & $ 139 $ & $ 57 $ & $ 4.8550 $ & $ 4.9084 $ & $ 8.378025 $ & $ 8.370554 $ & $ 1.899624618661961 $ \\
Ce & $ 140 $ & $ 58 $ & $ 4.8771 $ & $ 4.9193 $ & $ 8.376317 $ & $ 8.363506 $ & $ 1.898712581120185 $ \\
Pr & $ 141 $ & $ 59 $ & $ 4.8919 $ & $ 4.9301 $ & $ 8.353992 $ & $ 8.356420 $ & $ 1.897809775480838 $ \\
Ce & $ 142 $ & $ 58 $ & $ 4.9063 $ & $ 4.9410 $ & $ 8.347071 $ & $ 8.349298 $ & $ 1.896916042744723 $ \\
Nd & $ 142 $ & $ 60 $ & $ 4.9123 $ & $ 4.9410 $ & $ 8.346030 $ & $ 8.349298 $ & $ 1.896916042744723 $ \\
Nd & $ 143 $ & $ 60 $ & $ 4.9254 $ & $ 4.9517 $ & $ 8.330488 $ & $ 8.342138 $ & $ 1.896031227758121 $ \\
Sm & $ 144 $ & $ 62 $ & $ 4.9524 $ & $ 4.9624 $ & $ 8.303679 $ & $ 8.334943 $ & $ 1.895155179093460 $ \\
Nd & $ 145 $ & $ 60 $ & $ 4.9535 $ & $ 4.9731 $ & $ 8.309187 $ & $ 8.327712 $ & $ 1.894287748934647 $ \\
Nd & $ 146 $ & $ 60 $ & $ 4.9696 $ & $ 4.9837 $ & $ 8.304092 $ & $ 8.320446 $ & $ 1.893428792966205 $ \\
Nd & $ 148 $ & $ 60 $ & $ 4.9999 $ & $ 5.0048 $ & $ 8.277177 $ & $ 8.305813 $ & $ 1.891735743208532 $ \\
Sm & $ 149 $ & $ 62 $ & $ 5.0134 $ & $ 5.0153 $ & $ 8.263466 $ & $ 8.298446 $ & $ 1.890901377354422 $ \\
Sm & $ 150 $ & $ 62 $ & $ 5.0387 $ & $ 5.0258 $ & $ 8.261621 $ & $ 8.291047 $ & $ 1.890074941367280 $ \\
Sm & $ 152 $ & $ 62 $ & $ 5.0819 $ & $ 5.0465 $ & $ 8.244061 $ & $ 8.276153 $ & $ 1.888445348587004 $ \\
Eu & $ 153 $ & $ 63 $ & $ 5.1115 $ & $ 5.0568 $ & $ 8.228699 $ & $ 8.268659 $ & $ 1.887641943800672 $ \\
Sm & $ 154 $ & $ 62 $ & $ 5.1053 $ & $ 5.0671 $ & $ 8.226835 $ & $ 8.261135 $ & $ 1.886845972727570 $ \\
Gb & $ 154 $ & $ 64 $ & $ 5.1223 $ & $ 5.0671 $ & $ 8.224796 $ & $ 8.261135 $ & $ 1.886845972727570 $ \\
Gb & $ 155 $ & $ 64 $ & $ 5.1319 $ & $ 5.0773 $ & $ 8.213251 $ & $ 8.253581 $ & $ 1.886057318209837 $ \\
Dy & $ 156 $ & $ 66 $ & $ 5.1622 $ & $ 5.0875 $ & $ 8.192433 $ & $ 8.245997 $ & $ 1.885275865684704 $ \\
Gb & $ 156 $ & $ 64 $ & $ 5.1420 $ & $ 5.0875 $ & $ 8.215322 $ & $ 8.245997 $ & $ 1.885275865684704 $ \\
Gb & $ 157 $ & $ 64 $ & $ 5.1449 $ & $ 5.0977 $ & $ 8.203504 $ & $ 8.238385 $ & $ 1.884501503110640 $ \\
Gb & $ 158 $ & $ 64 $ & $ 5.1569 $ & $ 5.1078 $ & $ 8.201819 $ & $ 8.230743 $ & $ 1.883734120896110 $ \\
Dy & $ 158 $ & $ 66 $ & $ 5.1815 $ & $ 5.1078 $ & $ 8.190130 $ & $ 8.230743 $ & $ 1.883734120896110 $ \\
Tb & $ 159 $ & $ 65 $ & $ 5.0600 $ & $ 5.1178 $ & $ 8.188800 $ & $ 8.223074 $ & $ 1.882973611830810 $ \\
Dy & $ 160 $ & $ 66 $ & $ 5.1951 $ & $ 5.1279 $ & $ 8.184054 $ & $ 8.215377 $ & $ 1.882219871019077 $ \\
Gd & $ 160 $ & $ 64 $ & $ 5.1734 $ & $ 5.1279 $ & $ 8.183014 $ & $ 8.215377 $ & $ 1.882219871019077 $ \\
Dy & $ 161 $ & $ 66 $ & $ 5.1962 $ & $ 5.1379 $ & $ 8.173310 $ & $ 8.207652 $ & $ 1.881472795815909 $ \\
Er & $ 162 $ & $ 68 $ & $ 5.2246 $ & $ 5.1478 $ & $ 8.152397 $ & $ 8.199901 $ & $ 1.880732285764774 $ \\
Dy & $ 162 $ & $ 66 $ & $ 5.2074 $ & $ 5.1478 $ & $ 8.173457 $ & $ 8.199901 $ & $ 1.880732285764774 $ \\
Dy & $ 163 $ & $ 66 $ & $ 5.2099 $ & $ 5.1577 $ & $ 8.161785 $ & $ 8.192123 $ & $ 1.879998242537775 $ \\
Dy & $ 164 $ & $ 66 $ & $ 5.2218 $ & $ 5.1676 $ & $ 8.158714 $ & $ 8.184320 $ & $ 1.879270569877631 $ \\
Er & $ 164 $ & $ 68 $ & $ 5.2389 $ & $ 5.1676 $ & $ 8.149020 $ & $ 8.184320 $ & $ 1.879270569877631 $ \\
Ho & $ 165 $ & $ 67 $ & $ 5.2022 $ & $ 5.1774 $ & $ 8.146964 $ & $ 8.176490 $ & $ 1.878549173541836 $ \\
Er & $ 166 $ & $ 68 $ & $ 5.2516 $ & $ 5.1872 $ & $ 8.141959 $ & $ 8.168635 $ & $ 1.877833961248146 $ \\
Er & $ 167 $ & $ 68 $ & $ 5.2560 $ & $ 5.1970 $ & $ 8.131746 $ & $ 8.160756 $ & $ 1.877124842622655 $ \\
Er & $ 168 $ & $ 68 $ & $ 5.2644 $ & $ 5.2067 $ & $ 8.129601 $ & $ 8.152852 $ & $ 1.876421729148642 $ \\
Yb & $ 168 $ & $ 70 $ & $ 5.2702 $ & $ 5.2067 $ & $ 8.111898 $ & $ 8.152852 $ & $ 1.876421729148642 $ \\
Tm & $ 169 $ & $ 69 $ & $ 5.2256 $ & $ 5.2164 $ & $ 8.114473 $ & $ 8.144923 $ & $ 1.875724534117927 $ \\
Er & $ 170 $ & $ 68 $ & $ 5.2789 $ & $ 5.2260 $ & $ 8.111959 $ & $ 8.136971 $ & $ 1.875033172583114 $ \\
Yb & $ 170 $ & $ 70 $ & $ 5.2853 $ & $ 5.2260 $ & $ 8.106609 $ & $ 8.136971 $ & $ 1.875033172583114 $ \\
Yb & $ 171 $ & $ 70 $ & $ 5.2906 $ & $ 5.2356 $ & $ 8.097882 $ & $ 8.128995 $ & $ 1.874347561311723 $ \\
Yb & $ 172 $ & $ 70 $ & $ 5.2995 $ & $ 5.2452 $ & $ 8.097429 $ & $ 8.120996 $ & $ 1.873667618741784 $ \\
Yb & $ 173 $ & $ 70 $ & $ 5.3046 $ & $ 5.2547 $ & $ 8.087427 $ & $ 8.112975 $ & $ 1.872993264938715 $ \\
Yb & $ 174 $ & $ 70 $ & $ 5.3108 $ & $ 5.2642 $ & $ 8.083847 $ & $ 8.104930 $ & $ 1.872324421553529 $ \\
Lu & $ 175 $ & $ 71 $ & $ 5.3700 $ & $ 5.2737 $ & $ 8.069140 $ & $ 8.096864 $ & $ 1.871661011782399 $ \\
Hf & $ 176 $ & $ 72 $ & $ 5.3286 $ & $ 5.2831 $ & $ 8.061359 $ & $ 8.088776 $ & $ 1.871002960327636 $ \\
Yb & $ 176 $ & $ 70 $ & $ 5.3215 $ & $ 5.2831 $ & $ 8.064085 $ & $ 8.088776 $ & $ 1.871002960327636 $ \\
Hf & $ 177 $ & $ 72 $ & $ 5.3309 $ & $ 5.2925 $ & $ 8.051835 $ & $ 8.080666 $ & $ 1.870350193359751 $ \\
Hf & $ 178 $ & $ 72 $ & $ 5.3371 $ & $ 5.3019 $ & $ 8.049442 $ & $ 8.072535 $ & $ 1.869702638480259 $ \\
Hf & $ 179 $ & $ 72 $ & $ 5.3408 $ & $ 5.3113 $ & $ 8.038546 $ & $ 8.064383 $ & $ 1.869060224686682 $ \\
Hf & $ 180 $ & $ 72 $ & $ 5.3470 $ & $ 5.3206 $ & $ 8.034930 $ & $ 8.056210 $ & $ 1.868422882337254 $ \\
Ta & $ 181 $ & $ 73 $ & $ 5.3507 $ & $ 5.3298 $ & $ 8.023400 $ & $ 8.048017 $ & $ 1.867790543118027 $ \\
W & $ 182 $ & $ 74 $ & $ 5.3559 $ & $ 5.3391 $ & $ 8.018308 $ & $ 8.039804 $ & $ 1.867163140009889 $ \\
W & $ 183 $ & $ 74 $ & $ 5.3611 $ & $ 5.3483 $ & $ 8.008322 $ & $ 8.031571 $ & $ 1.866540607257118 $ \\
W & $ 184 $ & $ 74 $ & $ 5.3658 $ & $ 5.3574 $ & $ 8.005077 $ & $ 8.023318 $ & $ 1.865922880336927 $ \\
Os & $ 184 $ & $ 76 $ & $ 5.3823 $ & $ 5.3574 $ & $ 7.988677 $ & $ 8.023318 $ & $ 1.865922880336927 $ \\
Re & $ 185 $ & $ 75 $ & $ 5.3596 $ & $ 5.3666 $ & $ 7.991009 $ & $ 8.015046 $ & $ 1.865309895929486 $ \\
W & $ 186 $ & $ 74 $ & $ 5.3743 $ & $ 5.3757 $ & $ 7.988601 $ & $ 8.006755 $ & $ 1.864701591889463 $ \\
Os & $ 187 $ & $ 76 $ & $ 5.3933 $ & $ 5.3848 $ & $ 7.973780 $ & $ 7.998445 $ & $ 1.864097907217535 $ \\
Os & $ 188 $ & $ 76 $ & $ 5.3993 $ & $ 5.3938 $ & $ 7.973864 $ & $ 7.990117 $ & $ 1.863498782033505 $ \\
Os & $ 189 $ & $ 76 $ & $ 5.4016 $ & $ 5.4028 $ & $ 7.963002 $ & $ 7.981770 $ & $ 1.862904157549983 $ \\
Os & $ 190 $ & $ 76 $ & $ 5.4062 $ & $ 5.4118 $ & $ 7.962104 $ & $ 7.973405 $ & $ 1.862313976046459 $ \\
Ir & $ 191 $ & $ 77 $ & $ 5.3968 $ & $ 5.4208 $ & $ 7.948113 $ & $ 7.965023 $ & $ 1.861728180844594 $ \\
Os & $ 192 $ & $ 76 $ & $ 5.4126 $ & $ 5.4297 $ & $ 7.948525 $ & $ 7.956622 $ & $ 1.861146716284074 $ \\
Pt & $ 192 $ & $ 78 $ & $ 5.4169 $ & $ 5.4297 $ & $ 7.942491 $ & $ 7.956622 $ & $ 1.861146716284074 $ \\
Ir & $ 193 $ & $ 77 $ & $ 5.4032 $ & $ 5.4386 $ & $ 7.938133 $ & $ 7.948205 $ & $ 1.860569527698860 $ \\
Pt & $ 194 $ & $ 78 $ & $ 5.4236 $ & $ 5.4475 $ & $ 7.935941 $ & $ 7.939770 $ & $ 1.859996561394610 $ \\
Pt & $ 195 $ & $ 78 $ & $ 5.4270 $ & $ 5.4563 $ & $ 7.926552 $ & $ 7.931318 $ & $ 1.859427764626352 $ \\
Pt & $ 196 $ & $ 78 $ & $ 5.4307 $ & $ 5.4651 $ & $ 7.926529 $ & $ 7.922849 $ & $ 1.858863085576737 $ \\
Hg & $ 196 $ & $ 80 $ & $ 5.4385 $ & $ 5.4651 $ & $ 7.914369 $ & $ 7.922849 $ & $ 1.858863085576737 $ \\
Au & $ 197 $ & $ 79 $ & $ 5.4371 $ & $ 5.4739 $ & $ 7.915654 $ & $ 7.914364 $ & $ 1.858302473335395 $ \\
Pt & $ 198 $ & $ 78 $ & $ 5.4383 $ & $ 5.4827 $ & $ 7.914150 $ & $ 7.905863 $ & $ 1.857745877878276 $ \\
Hg & $ 198 $ & $ 80 $ & $ 5.4463 $ & $ 5.4827 $ & $ 7.911552 $ & $ 7.905863 $ & $ 1.857745877878276 $ \\
Hg & $ 199 $ & $ 80 $ & $ 5.4474 $ & $ 5.4914 $ & $ 7.905279 $ & $ 7.897345 $ & $ 1.857193250047730 $ \\
Hg & $ 200 $ & $ 80 $ & $ 5.4551 $ & $ 5.5001 $ & $ 7.905895 $ & $ 7.888812 $ & $ 1.856644541533518 $ \\
Hg & $ 201 $ & $ 80 $ & $ 5.4581 $ & $ 5.5088 $ & $ 7.897560 $ & $ 7.880263 $ & $ 1.856099704853727 $ \\
Hg & $ 202 $ & $ 80 $ & $ 5.4648 $ & $ 5.5174 $ & $ 7.896850 $ & $ 7.871698 $ & $ 1.855558693336665 $ \\
Tl & $ 203 $ & $ 81 $ & $ 5.4666 $ & $ 5.5260 $ & $ 7.886053 $ & $ 7.863118 $ & $ 1.855021461102933 $ \\
Hg & $ 204 $ & $ 80 $ & $ 5.4744 $ & $ 5.5346 $ & $ 7.885545 $ & $ 7.854523 $ & $ 1.854487963048309 $ \\
Pb & $ 204 $ & $ 82 $ & $ 5.4803 $ & $ 5.5346 $ & $ 7.879932 $ & $ 7.854523 $ & $ 1.854487963048309 $ \\
Tl & $ 205 $ & $ 81 $ & $ 5.4759 $ & $ 5.5432 $ & $ 7.878394 $ & $ 7.845912 $ & $ 1.853958154826700 $ \\
Pb & $ 206 $ & $ 82 $ & $ 5.4902 $ & $ 5.5517 $ & $ 7.875362 $ & $ 7.837287 $ & $ 1.853431992833947 $ \\
Pb & $ 207 $ & $ 82 $ & $ 5.4943 $ & $ 5.5602 $ & $ 7.869866 $ & $ 7.828648 $ & $ 1.852909434191629 $ \\
Pb & $ 208 $ & $ 82 $ & $ 5.5012 $ & $ 5.5687 $ & $ 7.867453 $ & $ 7.819994 $ & $ 1.852390436731829 $ \\
Bi & $ 209 $ & $ 83 $ & $ 5.5211 $ & $ 5.5772 $ & $ 7.847987 $ & $ 7.811325 $ & $ 1.851874958981518 $ \\
Ra & $ 226 $ & $ 88 $ & $ 5.7211 $ & $ 5.7172 $ & $ 7.661962 $ & $ 7.661932 $ & $ 1.843613253087140 $ \\
Th & $ 229 $ & $ 90 $ & $ 5.7557 $ & $ 5.7412 $ & $ 7.634650 $ & $ 7.635200 $ & $ 1.842246251422184 $ \\
Th & $ 230 $ & $ 90 $ & $ 5.7670 $ & $ 5.7491 $ & $ 7.630996 $ & $ 7.626267 $ & $ 1.841796196593272 $ \\
Th & $ 232 $ & $ 90 $ & $ 5.7848 $ & $ 5.7650 $ & $ 7.615033 $ & $ 7.608367 $ & $ 1.840904328898081 $ \\
U & $ 233 $ & $ 92 $ & $ 5.8203 $ & $ 5.7729 $ & $ 7.603956 $ & $ 7.599401 $ & $ 1.840462458679807 $ \\
U & $ 234 $ & $ 92 $ & $ 5.8291 $ & $ 5.7807 $ & $ 7.600715 $ & $ 7.590424 $ & $ 1.840023260016007 $ \\
U & $ 235 $ & $ 92 $ & $ 5.8337 $ & $ 5.7886 $ & $ 7.590914 $ & $ 7.581436 $ & $ 1.839586705273448 $ \\
U & $ 236 $ & $ 92 $ & $ 5.8431 $ & $ 5.7964 $ & $ 7.586484 $ & $ 7.572438 $ & $ 1.839152767221810 $ \\
U & $ 238 $ & $ 92 $ & $ 5.8571 $ & $ 5.8120 $ & $ 7.570125 $ & $ 7.554410 $ & $ 1.838292634239833 $ \\
Pu & $ 239 $ & $ 94 $ & $ 5.8601 $ & $ 5.8198 $ & $ 7.560318 $ & $ 7.545380 $ & $ 1.837866386796573 $ \\
Pu & $ 240 $ & $ 94 $ & $ 5.8701 $ & $ 5.8275 $ & $ 7.556042 $ & $ 7.536341 $ & $ 1.837442651004169 $ \\
\end{longtable}
\end{center}

\end{document}